\def\documenttype{arxiv} % TRANSITION BETWEEN arxiv or mitform

\def\arxiv{arxiv}

\ifx\documenttype\arxiv
    \documentclass{article}
    \usepackage{style/arxiv}
    \usepackage{amsmath}
    \usepackage{natbib}
    \usepackage[usenames, dvipsnames, hyperref]{xcolor}
    \definecolor{MyBlue}{RGB}{0, 20, 115}
    \usepackage[colorlinks=true, citecolor=MyBlue, linkcolor=MyBlue, urlcolor=MyBlue, breaklinks=true]{hyperref}
    \usepackage[capitalize]{cleveref}
    \usepackage[utf8]{inputenc}
    \usepackage[T1]{fontenc}
\fi

\usepackage[utf8]{inputenc} % allow utf-8 input
\usepackage[T1]{fontenc}    % use 8-bit T1 fonts
\usepackage{nicefrac}       % compact symbols for 1/2, etc.
\usepackage{microtype}      % microtypography
\usepackage{textcomp}

\usepackage{paracol}
\usepackage{pifont}

\usepackage{verbatim}
\usepackage{xspace}
\usepackage{setspace}

\usepackage{caption}
\usepackage{booktabs}
\usepackage{varwidth}
\usepackage{subcaption}
\usepackage{fancyhdr}
\usepackage{multirow}
\usepackage{enumitem}
\setlist{nosep, leftmargin=14pt}
\usepackage{url}      
\usepackage{color,soul}      
\usepackage{adjustbox}

\usepackage{graphicx}
\usepackage[misc]{ifsym}
\usepackage{amsmath}
\usepackage{amssymb}
\usepackage{amsfonts} 
\usepackage{bm}
\usepackage{dsfont}
\usepackage{mathtools}
\usepackage{siunitx}

\usepackage{pgf}
\usepackage{pgfplots}
\usepackage{pgffor}
\pgfplotsset{compat=1.18}

\usepackage{forloop}% http://ctan.org/pkg/forloop
\newcounter{loopcntr}

\crefname{equation}{Eq.}{Eqs.}
\crefname{figure}{Fig.}{Figs.}
\crefname{section}{Sec.}{Secs.}
\crefname{table}{Tab.}{Tabs.}
\crefname{appendix}{Appx.}{Appx.}
\Crefname{equation}{Equation}{Equations}
\Crefname{figure}{Figure}{Figures}
\Crefname{section}{Section}{Sections}
\Crefname{table}{Table}{Tables}
\Crefname{appendix}{Appendix}{Appendices}

\graphicspath{ {./images/} }

\newcommand{\rom}[1]{\uppercase\expandafter{\romannumeral #1\relax}}

\newcommand{\mnm}[1]{\mbox{#1 nm}\xspace}
\newcommand{\mmu}[1]{\mbox{#1 {\textmu}m}\xspace}
\newcommand{\msmu}[1]{\mbox{#1 {\textmu}m²}\xspace}

\newcommand{\mmm}[1]{\mbox{#1 mm}\xspace}
\newcommand{\mpx}[1]{\mbox{#1 pixels}\xspace}
\newcommand{\mdg}[1]{\mbox{#1\textdegree}\xspace}

\newcommand{\mpli}{\mbox{3D-PLI}\xspace}
\newcommand{\crv}{Cresyl violet\xspace}

\newcommand{\gram}{Gram\xspace}
\newcommand{\gramreg}{Gram+Reg\xspace}
\newcommand{\gan}{GAN\xspace}
\newcommand{\ganreg}{GAN+Reg\xspace}

\DeclareMathOperator*{\argmax}{arg\,max}
\DeclareMathOperator*{\argmin}{arg\,min}

\ifx\documenttype\arxiv
  \renewcommand{\cite}[1]{\citep{#1}}
  \newcommand{\prefixcite}[2]{(#1, \citealt{#2})}
\fi

\title{From Fibers to Cells: Fourier-Based Registration Enables Virtual Cresyl Violet Staining From 3D Polarized Light Imaging}

\def\institutelist#1{
  #1$^{1}$ Institute of Neuroscience and Medicine (INM-1), Research Centre Jülich, Jülich, Germany \\
  #1$^{2}$ Helmholtz AI, Research Centre Jülich, Germany \\
  #1$^{3}$ Cécile \& Oskar Vogt Institute of Brain Research, Medical Faculty and University Hospital Düsseldorf, \\ #1 Heinrich Heine University Düsseldorf, Germany \\
  #1$^{4}$ Department of Physics, University of Wuppertal, Germany \\
  #1$^{5}$ Institute of Computer Science, Faculty of Mathematics and Natural Sciences, \\ #1 Heinrich Heine University Düsseldorf, Düsseldorf, Germany
}
        
\def\authornamearxiv{
  \bf{
    Alexander Oberstrass$^{1,2,}$\thanks{Equal contribution.}~\;
    Esteban Vaca$^{1,2,}$\footnotemark[1]~\;
    Eric Upschulte$^{1,2,3}$\;
    Meiqi Niu$^{1}$
  }
  \\
  \bf{
    Nicola Palomero-Gallagher$^{1,3}$\;
    David Graessel$^{1}$\;
    Christian Schiffer$^{1,2}$
  }
  \\
  \bf{
    Markus Axer$^{1,4}$\;
    Katrin Amunts$^{1,3}$\;
    Timo Dickscheid$^{1,2,5}$
  }
  \\[5mm]
  \institutelist{\normalsize}
}

\ifx\documenttype\arxiv
  \author{\authornamearxiv} 
\fi

\begin{document}

  \maketitle

  \begin{abstract}
  Comprehensive assessment of the various aspects of the brain's microstructure requires the use of complementary imaging techniques.
  This includes measuring the spatial distribution of cell bodies (cytoarchitecture) and nerve fibers (myeloarchitecture).
  The gold standard for cytoarchitectonic analysis is light microscopic imaging of cell-body stained tissue sections.
  To reveal the 3D orientations of nerve fibers, 3D Polarized Light Imaging (\mpli) has been introduced as a reliable technique providing a resolution in the micrometer range while allowing processing of series of complete brain sections.
  \mpli acquisition is label-free and allows subsequent staining of sections after \mpli measurement.
  By post-staining for cell bodies, a direct link between fiber- and cytoarchitecture can potentially be established in the same section.
  However, inevitable distortions introduced during the staining process make a costly nonlinear and cross-modal registration necessary in order to study the detailed relationships between cells and fibers in the images.
  In addition, the complexity of processing histological sections for post-staining only allows for a limited number of such samples.
  In this work, we take advantage of deep learning methods for image-to-image translation to generate a virtual staining of \mpli that is spatially aligned at the cellular level.
  We use a supervised setting, building on a unique dataset of brain sections, to which \crv staining has been applied after \mpli measurement.
  To ensure high correspondence between both modalities, we address the misalignment of training data using Fourier-based registration.
  In this way, registration can be efficiently calculated during training for local image patches of target and predicted staining.
  We demonstrate that the proposed method can predict a \crv staining from \mpli, resulting in a virtual staining that exhibits plausible patterns of cell organization in gray matter, with larger cell bodies being localized at their expected positions.

\end{abstract}

  % Introduction
  \section{Introduction}
\label{sec:intro}

To understand the organizational principles of the brain, complementary imaging techniques are used to highlight different aspects of brain architecture.
Two important aspects of the microstructural organization are fiber- and cytoarchitecture \cite{nieuwenhuys2013, amunts2015}.
While cytoarchitecture encompasses the spatial distribution and shape of cell bodies in the cerebral cortex and subcortical nuclei, fiber architecture refers to the course and composition of nerve fibers.
However, cytoarchitecture and fiber-architecture are usually studied using different staining protocols, applied in different sections.
Only a few protocols are available to combine cyto- and fiber staining in a single protocol, e.g., Luxol fast blue \cite{kluver1953}, Bielschowsky \cite{bielschowsky1904} or the triple staining by Novotny \cite{novotny1977}.
While they allow visualizing cell bodies and fibers in one and the same section, they lack information about 3D fiber orientations.
As a result, they do not support the tracing of axons and fiber bundles over long distances.

3D-Polarized Light Imaging (\mpli) addresses this limitation.
It is a microscopic imaging technique for evaluating the three-dimensional orientation of myelinated nerve fibers in entire, unstained histological brain sections  \cite{axer2022, axer2011, axer2011a}.
The technique can achieve an in-plane resolution of \mmu{1.3}, capturing structures at the level of individual fibers and small fiber bundles.
\mpli has been used to gain insights into the architecture of nerve fibers in different brain regions, such as the human hippocampus \cite{zeineh2017}, the sagittal stratum \cite{caspers2022} and the vervet monkey visual system \cite{takemura2020}.
In addition, 3D-PLI has been used to validate fiber tractography algorithms and the interpretation of DW-MRI \cite{caspers2019}.
\mpli potentially allows joint imaging of fiber tracts and neuronal cell bodies \cite{zeineh2017} due to diffraction patterns, differences in the density of birefringent material, and locally variable attenuation.
However, this possibility has not yet been validated.

Cytoarchitecture can be studied in histological sections of postmortem brains with \crv.
The staining provides contrast due to staining of the rough endoplasmic reticulum.
This allows to study  cell shape, density and distribution, which vary between brain regions.
Due to its high spatial resolution, microscopic analysis of histological sections is considered the gold standard to verify structural parcellations \cite{amunts2015}.
Recent advances in high-throughput scanning, data processing algorithms, and computational capacities have enabled the creation of 3D human brain atlases based on cytoarchitecture, such as BigBrain \cite{amunts2013}, the Allen Adult Human Brain Atlas \prefixcite{AAHA}{ding2016}, and the Julich Brain probabilistic atlas \cite{amunts2020}.

Since \mpli relies solely on optical properties of the tissue, it is label-free and can be combined with a staining of the same tissue after its measurement.
Post-staining, e.g. with \crv, enables a complementary visualization of neuronal cell bodies, potentially establishing a direct link between cytoarchitecture with 3D fiber-architecture.
This requires, however, a complex histological processing, which limits the number of available samples, increases the risk of tissue damage and may lead to deformations of the section.
To correct for deformation and artifacts in the images of the two modalities requires a nonlinear registration step.
However, \mpli and \crv stained tissue share only a few automatically identifiable cross-modal registration landmarks, such as blood vessels or morphological landmarks.
Therefore, post-staining of sections imaged with \mpli is feasible but technically challenging.
Since it does not scale efficiently to larger datasets, it is not applicable to whole-brain stacks involving thousands of sections.

\begin{figure*}[]
    \centering
    \includegraphics[width=\textwidth]{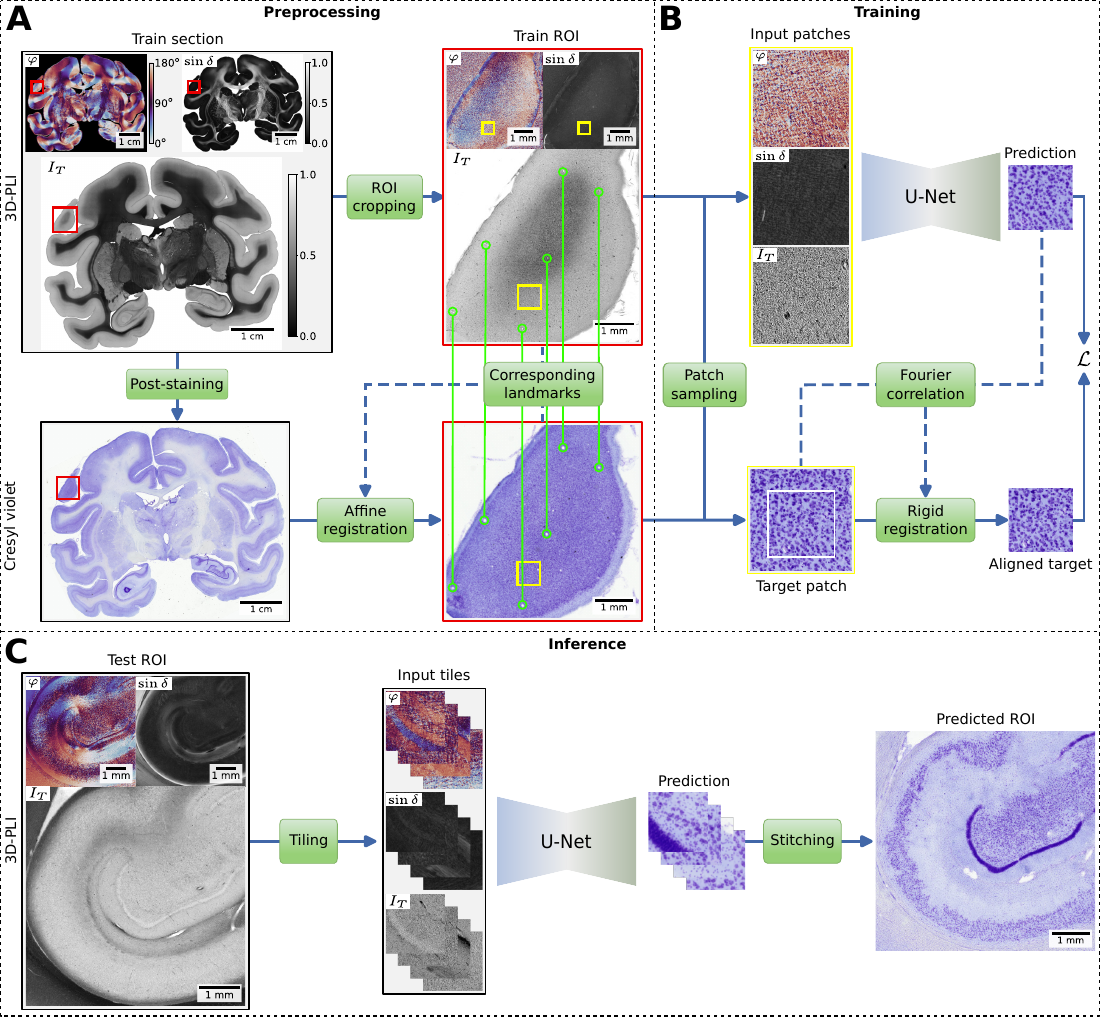}
    \caption{
      The proposed virtual staining workflow.
      \textbf{(A)} Preprocessing of \mpli sections that were post-stained with \crv.
      As paired training data, regions of interest (ROIs) are manually cropped (red boxes) and affine registered using large blood vessels as landmarks (green marker).
      Background pixels in train sections are masked, and retardation values are scaled using gamma correction for visualization purposes.
      \textbf{(B)} Training of a U-Net model using patches, extracted from same random locations (yellow boxes) in \mpli modalities direction $\varphi$, retardation $\sin \delta$, transmittance $T_T$ and the \crv staining.
      \mpli patches are used as input to the model to predict a virtual \crv staining.
      The \crv patch acts as target and is rigidly aligned with the prediction during the training procedure.
      The alignment is performed by our proposed online registration head using Fourier-based correlation of pixels. 
      A loss $\mathcal{L}$ is computed between aligned target and prediction.
      \textbf{(C)} Inference using the trained U-Net model to virtually stain unseen sections or ROIs.
      Inputs are divided into overlapping tiles, which are processed independently by the U-Net model.
      The predictions are then stitched back together to form the complete virtual staining.
    }
    \label{fig:total_overview}
  \end{figure*}

Therefore, we aim to train a deep neural network model to perform image-to-image translation from \mpli to a \crv staining.
Such an approach is often denoted as \textit{virtual histological staining}, which refers to computational methods that generate color-coded images of biological tissue without the need for traditional staining techniques \cite{latonen2024}.
The methods instead utilize optical properties of the tissue, such as birefringence, autofluorescence, scattering, or absorption to create images that emulate the appearance of stained tissue.
A virtual \crv staining spatially aligned with \mpli would allow the use of established tools for cytoarchitectonic analysis, such as automatic cell instance segmentation \cite{upschulte2022}, directly on \mpli data.
Furthermore, identification of cell bodies would provide detailed registration landmarks for cross-modal registration \cite{ounkomol2018}, thereby offering the opportunity to perform joint acquisition of aligned fiber and cytoarchitecture at a larger scale than possible today.
A virtual staining could, in principle, be applied to the whole brain.
However, white matter regions remain challenging due to the predominance of glial cells, mainly oligodendrocytes forming the myelin sheath, which are not distinguishable from nerve fibers in the \mpli signal. 
Therefore, in the present study, we focus our analysis on gray matter regions.

One of the earliest applications of label-free imaging for visual staining was Quantitative Phase Imaging \prefixcite{QPI}{curl2004}.
QPI measures the phase shift of light passing through a sample, producing high-resolution images that reveal the optical properties of the tissue. 
It was used to generate images of collagen fibers, red blood cells, and other tissue structures without staining \cite{curl2004,park2018}.
Later, machine-learning algorithms, especially generative models, have been trained to recognize and virtually stain different tissue structures in unstained images by performing image-to-image translation.
They have successfully generated color-coded images of tissue that replicate the appearance of histological stainings, such as a virtual hematoxylin and eosin (H\&E), Masson's trichrome, and Jones' stain from QPI of label-free tissue \cite{rivenson2019}, a transformation of H\&E stained tissue into Masson's trichrome, periodic acid-Schiff (PAS), or Jones' stain \cite{dehaan2021,yang2022}.
However, these stains are not very good at distinguishing the different components of the nervous tissue.

Machine learning algorithms were also used to predict \mbox{fluorescence-labeled} images from \mbox{transmitted-light} \mbox{z-stacks} \cite{ounkomol2018, christiansen2018, cross-zamirski2022} or 3D fluorescence structures and a FluoroMyelin stain from bright-field and polarization images of brain slices \cite{guo2020}.
The methods typically use cross-entropy \cite{christiansen2018}, mean absolute (L1) loss \cite{guo2020}, or style-related losses such as conditional generative adversarial network (GAN) loss \cite{rivenson2019, dehaan2021, cross-zamirski2022}.
While including a GAN objective encourages prediction of realistic-looking images, it has no clear mechanism to preserve content when conditioned on a particular input image, and thus may introduce artificial structures \cite{cohen2018}.
A combination with a pixel-wise reconstruction loss (e.g. L1 loss) mitigates this problem of GAN training and improves accuracy of predictions \cite{isola2017}.

Since methods using paired training data for supervised image-to-image translation typically produce more accurate predictions than unpaired methods \cite{zhu2017,latonen2024}, a pixel-accurate alignment of training data is desired.
This requires virtual staining methods to either perform a costly registration step or directly acquire paired images.
A paired acquisition with \mpli, however, is not feasible and a lack of structural overlap, such as a sufficient number of visible cell instances between the investigated modalities, makes pixel-accurate registration challenging.
Therefore, to alleviate the need for perfectly paired training data, we propose a supervised learning objective performing local online registration of training pairs combined with a translational-invariant style comparison.
This allows us to train the model on imperfectly registered image pairs with strong content preservation as in paired image-to-image translation, while enabling realistic prediction of subtle structures like cell bodies (\cref{fig:total_overview}).

The main contributions of our method are the following:

\begin{itemize}
    \item We apply the matching of Gram matrix representations as a \textit{texture sensitive style loss} for the virtual staining, as previously used for texture synthesis \cite{gatys2015a}.
    Since the computation of Gram matrices is translation invariant, it allows a direct comparison of image statistics between coarsely aligned training examples.
    It therefore improves the accuracy of predicted cell instances over commonly used GAN style loss.
    \item \textit{An online registration head} for improving registration accuracy of local image pairs after pre-alignment of larger tissue tiles during training.
    We consider Fourier-based registration methods, which can be computed efficiently in real-time on modern GPU hardware.
    \item \textit{An equivariance loss} to improve the accuracy of cell instance predictions by addressing the inherent agnosticism of loss computation to constant displacements through online registration.
\end{itemize}

  % Methods
  \section{Materials and Methods}
\label{sec:methods}

\subsection{Microscopic imaging of histological brain sections}
\label{sec:dataset}

We demonstrate the proposed virtual staining approach on a set of brain sections for which \crv staining has been performed after \mpli measurement.
The brain sample for this study was obtained from a healthy 2.4-year-old adult male vervet monkey \prefixcite{Wake Forest-ID 1818}{axer2020vervet,takemura2020} in accordance with the Wake Forest Institutional Animal Care and Use Committee (\mbox{IACUC \#A11-219}) and conforming the AVMA Guidelines for the Euthanasia of Animals.
To obtain an undistorted volumetric reference, a T2 weighted MRI was acquired in-vivo one day prior to sacrifice.
The brain was removed from the skull within 24 hours after flushing with phosphate-buffered saline and perfusion fixated with 4\% buffered paraformaldehyde.
It was stored for several weeks at -70\textdegree C in 20\% glycerin solution for cryoprotection, and then sliced coronally with \mmu{60} section thickness using a large-scale cryostat microtome (Polycut CM 3500, Leica Microsystems, Germany).
Blockface images of the frozen tissue block were taken with a CCD camera before cutting each brain section.
The images were reconstructed into a 3D blockface volume to provide an undistorted reference for section realignment \cite{schober2015}.

\subsubsection{Image acquisition}

\begin{figure*}[t]
  \centering
  \includegraphics[width=\textwidth]{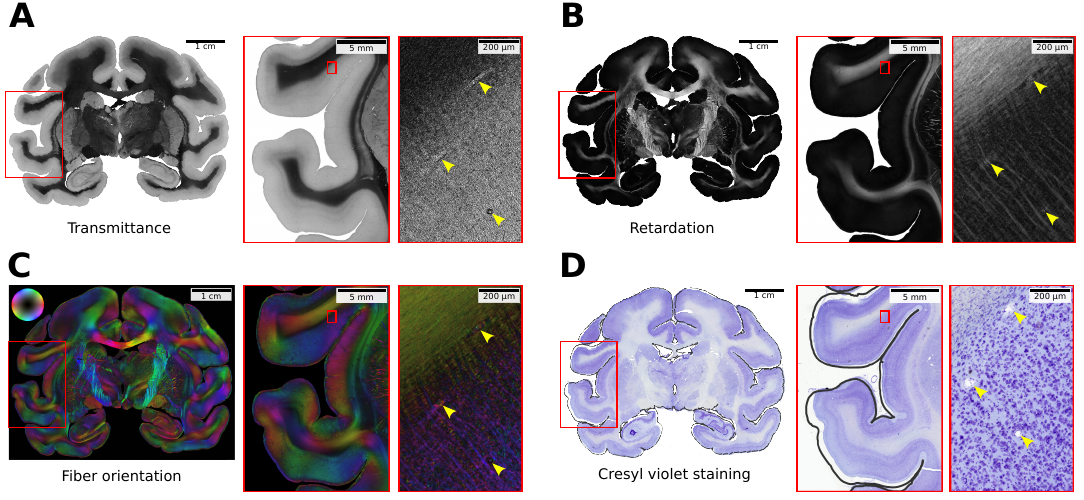}
  \caption{
    Data modalities and registration challenges for training section 544.
    \textbf{(A-C)} \mpli parameter maps: Transmittance,
    retardation and fiber orientation in HSV color space (hue: fiber direction; saturation/brightness: retardation).
    Background pixels are masked for visualization purposes only.
    \textbf{(D)} Affine registered \crv staining.
    The pial surface of the \mpli acquisition is shown as a contour plot in D for reference.
    Between both data acquisitions remains a nonlinear misalignment that cannot be resolved by a global affine transformation.
    At a local scale, the remaining misalignment is approximately linear.
    Yellow arrows indicate blood vessels that can be used as mutual registration landmarks for coarse alignment.
  }
  \label{fig:dataset}
\end{figure*}

For \mpli acquisition, brain sections were scanned using a polarizing microscope (LMP-1, Taorad, Germany) with \mmu{1.3} resolution \cite{axer2011a, axer2022}.
The focus level of the LMP-1 was manually adjusted to the center of the tissue for each section.
Inside the LMP-1 microscope, sections were placed on a specimen stage between a rotating linear and a circular polarizer on top of an incoherent light source with a wavelength of \mnm{550 $\pm$ 5}.
Images were taken by a CCD camera for nine equidistant rotation angles $\rho$ of the rotating linear polarizer, covering \mdg{180} of rotation.
At each pixel, the measured intensity of the images followed a sinusoidal profile as
\begin{equation}
  \label{eq:pli_formula}
  I_\rho = \frac{I_T}{2} \left(1 + \sin(2\rho - 2\varphi) \sin \delta \right).
\end{equation}
Using harmonic Fourier analysis, parameter maps of transmittance ($I_T$), retardation ($\sin \delta$), and fiber direction ($\varphi$) were obtained from the measurements, with an image size of approximately \mpx{34,000 $\times$ 44,000} per section, revealing their fine-grained nerve fiber architecture (\cref{fig:dataset}A-C).
Each parameter map was stored in a separate HDF5 file as uncompressed 32-bit floating-point single-channel image.

After \mpli acquisition, brain sections were washed, fixed and stained for cell bodies with \crv Nissl staining to reveal their cellular architecture.
Whole-slide flat scans (single-plane) were performed using a Huron TissueScope LE120 high-throughput scanner at \mmu{1} in-plane resolution (\cref{fig:dataset}D).
The resulting images were saved as RGB color images with eight bit color depth (pixel values ranging from 0 to 255) in uncompressed BigTIFF format.

\subsubsection{Optical effects of cell bodies on the 3D-PLI signal}

While \mpli was primarily developed to map fiber orientations, cell bodies contribute to the measured signal as well.
In the following, we summarize how absorption, diffraction, birefringence and scattering effects of cells are represented in \mpli parameter maps.

Previous work reported that larger cell bodies appear as dark spots in transmittance maps \cite{zeineh2017}, which encode light extinction caused by any material along the optical path.
However, in the present transmittance maps, cell bodies are not distinguished by higher absorption relative to the surrounding fiber architecture (\cref{fig:dataset}A).
This is likely because their membranes remained intact due to the short postmortem time before tissue fixation.

Diffraction significantly impacts the \mpli signal at the given wavelength and resolution.
In cortical regions, diffraction can cause pixels of transmittance maps inside the tissue to appear brighter than the background, particularly along sharp edges such as the walls of cell bodies and blood vessels.
The intensity of these diffraction patterns on the transmittance map depends on the level of the focal plane of the objective lens.

Another relevant effect is observed in retardation maps, which encode the average amount and orientation of birefringent material within each tissue voxel, primarily collagen and myelinated nerve fibers.
Since cell bodies contain significantly less birefringent material than surrounding fibers, their presence causes local attenuation of the retardation appearing as dark patches between cortical fibers \mbox{(\cref{fig:dataset}B)}.

Light scattering is ubiquitous in polarized light bright-field
transmission microscopy, but can generally be treated as random background noise.
It does not appear to systematically affect either transmittance or retardation, with the only known exception being increased scattering at steep fibers running approximately perpendicularly to the section plane and darkening the transmittance significantly.

\subsubsection{Tissue shrinkage estimation}
\label{sec:tissue_shrinkage_correction}
Deformations induced by histological processing include shrinkage or swelling of brain tissue.
To correct these effects, the extent of shrinkage can be estimated from the ratio between the histologically processed and true brain volume, which can be represented by the fresh weight of the whole brain with an estimated mean specific density \cite{amunts2005,amunts2013} or an MRI reference \cite{wagstyl2020}.
In this work, we perform 2D segmentation of cell bodies in \mpli parameter maps and measure their in-plane areas.
To ensure comparability of cell sizes across studies, we estimate 2D shrinkage factors using a postmortem MRI of the same brain as a reference.

We first estimate shrinkage in the \mpli acquisition by affine registration of the 3D reconstructed blockface volume to the MRI.
The linear part of the affine transformation in physical space has eigenvalues [0.984, 1.005, 1.015], indicating a global volume change of less than 1\% and no axis-specific systematic deviation.
In a second step, we calculate 2D shrinkage factors for each brain section as the quotient between the area occupied by tissue in the \mpli measurement and its corresponding blockface image.
For test section \#559, we estimate a global 2D shrinkage factor of 0.97, indicating a slight swelling of the \mpli measurement relative to its original area within the MRI.
This observation is consistent across all sections, with 2D shrinkage factors between 0.95 and 0.99.
We correct the in-plane sizes of segmented cell body areas in \mpli by applying the individual 2D shrinkage factor of each section, assuming an approximately uniform area change of cells and surrounding tissue.

\subsection{Initial cross-modality alignment}
\label{sec:cm_alignment}

After the subsequent processing of brain tissue, \crv images exhibit a deformation relative to the \mpli acquisition.
To align both modalities, an initial affine registration of whole brain sections is performed by manual identification of large blood vessels as landmarks visible in both modalities.

Performing the initial affine registration reveals remaining nonlinear deformations as shown in \cref{fig:dataset}D.
Since nonlinear deformations typically have low spatial frequencies, causing smooth, large-scale distortions, we expected near-linear deformations at smaller scales.
Therefore, performing an additional more local linear registration would lead to a better fit.
We subsequently crop square regions of interest (ROIs) with a size of \mpx{4,096} ($\sim$\mmm{5.3}) and without visible artifacts, covering distinct cellular architectures across the whole coronal plane.
For all ROIs we perform additional affine registration and make sure that transformed landmarks have a maximum distance of \mpx{70} (\mmu{91}) from their matches (\cref{fig:total_overview}A).
All ROIs are warped and resampled to \mmu{1.3} using linear interpolation to match the coordinate space of \mpli.
While this results in a loss of precision relative to the original resolution of \crv scans of \mmu{1}, matching the resolutions of both modalities facilitates subsequent processing and analysis steps.

\subsection{Fourier-based online registration of image patches}
\label{sec:fft_reg}

To correct  the remaining misalignment of \mpli and \crv after affine registration of ROIs at a finer local scale, we introduce an online registration head that performs cross-modality alignment during training based on model predictions of small image patches (\cref{fig:total_overview}B).
We assume that once the style transfer model has learned to reconstruct microscopic landmarks (e.g., individual cells or small blood vessels), such online registration will promote the learning of additional landmarks until the training can use pixel-aligned training examples.

The registration method performed during training needs to be computationally efficient, since training will require numerous registration iterations.
Conventional feature-based image registration methods are accurate and can model nonlinear deformations but are computationally expensive and sensitive to image degradation.
As we assume deformations to be approximately linear at a local scale, we take advantage of Fourier-based image correlations \cite{tong2019}, which can efficiently recover a translation between images in the frequency domain.

\subsubsection{Translational shift}
Fourier-based image correlation methods are able to retrieve a translational shift $(\Delta u,\Delta v)$ between image functions $f(u, v)$ and $g(u, v)$ defined for integer pixel coordinates $(u, v)$, such that $f(u, v) = g(u + \Delta u, v + \Delta v)$.
Both functions $f$ and $g$ represent images of equal height $H$ and width $W$ and are for now assumed to repeat periodically with a periodicity of $H$ and $W$, respectively.

A common approach to retrieve the translational shift between $f$ and $g$ is to use circular cross-correlation \cite{tong2019}, which can be efficiently computed in the frequency domain as
\begin{align}
  \begin{split}
    \text{CC°}[a, b] &= (f \star g)[a, b] \\
    &= \sum_{u, v} f(u - a, v - b)g(u, v) \\
    &= \left(\mathcal{F}^{-1}\left\{\overline{\mathcal{F}\{f\}} \mathcal{F}\{g\}\right\}\right)[a, b],
  \end{split}
	\label{eq:cross_corelation}
\end{align}
for all integer shifts $[a, b]$, where $\mathcal{F}$ denotes the Fourier transformation, $\mathcal{F}^{-1}$ its inverse, $\overline{\mathcal{F}\{f\}}$ its complex-conjugate Fourier coefficients, and where we sum over all pixel coordinates $(u, v)$.
The translational shift can subsequently be recovered by the location of the maximum value in $\text{CC°}$ as
\begin{equation}
  (\Delta u, \Delta v) = \argmax_{(a, b)} \text{CC°}[a,b].
  \label{eq:translation_shift}
\end{equation}
\cref{eq:cross_corelation} can be extended to the mean over squared distances between pixel values as
\begin{align}
  \begin{split}
    \text{MSE°}[a, b] &= \frac{1}{HW} \sum_{u, v} (f(u - a, v - b) - g(u, v))^2	\\
    &= \frac{- 2 (f \star g)[a, b] + \sum_{u, v} (f^2(u, v) + g^2(u, v))}{HW} .
  \end{split}
  \label{eq:fourier_mse_circular}
\end{align}
The translational shift $(\Delta u, \Delta v)$ can be recovered analog to \cref{eq:translation_shift} by computing the $\argmin$.
For periodic image functions $f$ and $g$, solutions of $\text{CC°}$ and $\text{MSE°}$ are identical as the sum over squared functions $f^2$ and $g^2$ is constant \cite{fienup1997}.

Since histological images are not periodic,
zero padding is applied to fill both images up to a shape of $(H_f + H_g - 1, W_f + W_g - 1)$, in order to break periodicity and allow processing of images with different heights $H_f$, $H_g$ and widths $W_f$, $W_g$. 
The zero-padded images are denoted by new image functions $f_0$ and $g_0$.
Furthermore, additional masks $M_f$ and $M_g$ are introduced, which have a value of one at all pixel coordinates within original height and width and zero elsewhere.
We reformulate \cref{eq:fourier_mse_circular} to a non-circular form as
\begin{equation}
  \text{MSE}[a, b] = \left( \frac{(f_0^2 \star M_g) -2 (f_0 \star g_0) + (M_f \star g_0^2)}{M_f \star M_g} \right)[a, b] ,
  \label{eq:fourier_mse}
\end{equation}
which can be efficiently computed by multiple applications of \cref{eq:cross_corelation}.
Here, the cross-correlations of $f_0^2$ and $g_0^2$ with masks $M_g$ and $M_f$, respectively, ensure that pixel values of the original unpadded images are not compared with zero padding values.
Furthermore, the score is divided by the correlation between $M_f$ and $M_g$ to account for the number of overlapping pixels between original unpadded images.
We apply the same division by $M_f \star M_g$ also for CC° to retrieve a non-circular variant called CC.
While the solutions for circular CC° and MSE° are identical, the solutions for non-circular CC and MSE differ, resulting in distinct registration metrics.

\subsubsection{Rotation and scale}\label{sub:rot_scale}
While relative scale and rotation between images can also be retrieved in the frequency domain \cite{sheng1989,reddy1996}, we expect only small relative rotation angles and minor scale variations due to the initial affine registration by matching blood vessels.
Therefore, we leverage the parallel processing capability of GPUs to perform an exhaustive search over a fixed set of rotation angles without scaling adjustments.
We calculate registration metrics for every translational shift and rotation and select the rotation angle and shift combination that yields a global optimum.

\begin{figure*}[t]
  \centering
  \includegraphics[width=\textwidth]{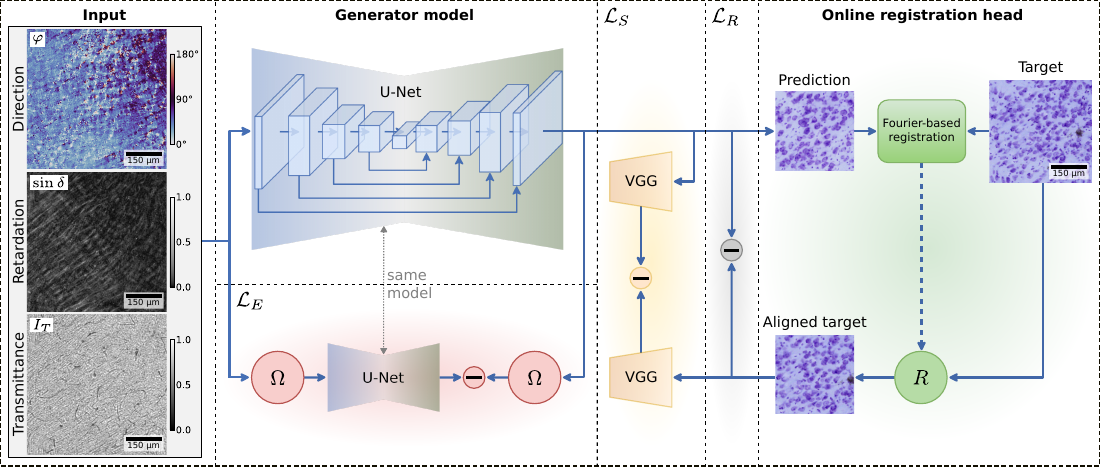}
  \caption{
    Illustration of the proposed virtual staining approach.
    Patches of \mpli parameter maps transmittance $I_T$, retardation $\sin \delta$ (scaled using gamma correction for visualization), and direction $\varphi$ are used as input to a 2D convolutional U-Net model as generator to predict a virtual \crv staining.
    An online registration head estimates a rigid transformation $R$ between a coarsely aligned \crv target patch and the prediction via Fourier-based registration.
    Transformation $R$ is used to align target and prediction at the patch level.
    We calculate three distinct loss components: $\mathcal{L}_R$, $\mathcal{L}_S$ and $\mathcal{L}_E$.
    Reconstruction loss $\mathcal{L}_R$ performs a pixel-wise comparison between prediction and aligned target.
    Style loss $\mathcal{L}_S$ compares feature maps of a VGG network encoder using Gram matrices to mimic the style of the target image.
    Equivariance loss $\mathcal{L}_E$ applies the same U-Net model a second time to a rotated version of the input by rotation $\Omega$.
    The output is compared with the prediction rotated by same rotation $\Omega$, which promotes stability and avoids learning a constant shift of pixels in the prediction.
  }
  \label{fig:main_scheme}
\end{figure*}

\subsection{Conditional generation of Cresyl violet staining}
\label{sec:generation_method}

Image-to-image translation refers to the process of generating images using a generator model $G$ conditioned on an input image $x$.
Predictions $G(x)$ of the model are compared with actual target images $y$.
In our case, we translate from the domain of \mbox{\mpli} images to \mbox{\crv}.
The translation is performed by a U-Net \mbox{\cite{ronneberger2015}} serving as the generator model, which forms the core of our image-to-image translation framework, as illustrated in \mbox{\cref{fig:main_scheme}}.
The three loss components used to train the U-Net are described below.

\subsubsection{Reconstruction Loss}
Similar to \cite{isola2017}, we use an $L_1$ loss between target $y$ and prediction $G(x)$ to encourage pixel correspondence in the \mbox{\textit{reconstruction loss}}
\begin{equation}
    \mathcal{L}_{R} = \mathbb{E}_{x,y} \left[\left\lVert R_{y,G(x)}(y) - G(x)\right\rVert_1 \right],
    \label{eq:loss_reconstruction}
\end{equation}
where $R_{y,G(x)}$ performs the proposed online registration to spatially align $y$ and $G(x)$ before loss calculation.
This enables utilization of imperfectly aligned training data to penalize any discrepancies in corresponding pixel values between $G(x)$ and $y$.
To disable online registration, \mbox{$R_{y,G(x)}$} can be replaced by the identity function.

\subsubsection{Style Loss}
We consider two alternative implementations of a style loss $\mathcal{L}_{S}$, both focussing on style preservation.
We refer to the first one as \textit{Gram loss}, and the second as \textit{GAN loss}.

For the Gram loss, we apply a texture-sensitive style loss proposed by \citet{gatys2015a} based on squared distances between Gram matrix representations of neural network features.
Given a pre-trained VGG encoder \cite{simonyan2015}, Gram matrix representations are computed from feature activations of its layers to characterize the texture of images at different complexities.
For each layer $l$, the encoder produces a different number of $N_l$ feature maps, each storing $K_l$ spatial entries (i.e., height $\times$ width).
Elements of the Gram matrix $\Gamma^l_{ij}$ at layer $l$ are computed as the inner product between the $i$-th and $j$-th feature map $F_i^l$ and $F_j^l$, where each map is flattened to a $K_l$-dimensional vector:
\begin{equation}
  \Gamma^l_{ij} = \sum^{K_l}_{k=1} F^l_{ik}F^l_{jk}.
\end{equation}
Since the Gram matrix computation captures global feature correlations rather than spatial locations, this allows a translation-invariant comparison of image statistics.
The style loss is computed over all $L$ layers of the VGG encoder, using Gram matrix representations $\Gamma^l$ for the online registered target and $\hat{\Gamma}^l$ for the prediction:
\begin{equation}
  \mathcal{L}_{S} = \mathbb{E}_{x,y} \left[ \sum^{L}_{l=1} \frac{1}{K_l^2 N_l^2} \sum^{N_l}_{i=1} \sum^{N_l}_{j=1} (\Gamma^l_{ij} - \hat{\Gamma}^l_{ij})^2 \right].
\end{equation}

For the GAN loss, we consider adversarial training \cite{goodfellow2014} to compare with previous work in virtual staining \cite{rivenson2019, cross-zamirski2022}.
We implement the GAN loss in the form of a Wasserstein GAN \cite{arjovsky2017a}.
In contrast to conditional GAN training \cite{mirza2014, isola2017}, we do not condition the discriminator on input images $x$, as this would cause the model to reproduce any misalignment in the training data.

\subsubsection{Equivariance Loss}
By registering target $y$ to generator prediction $G(x)$ before loss calculation, displacements of objects in $G(x)$ relative to $x$ are not captured by $\mathcal{L}_{R}$.
To prevent pixel shifts in $G(x)$, we enforce equivariance with respect to rotations through the \mbox{\textit{equivariance loss}}
\begin{equation}
    \mathcal{L}_{E} = \mathbb{E}_{x} \left[ \left\lVert \Omega(G(x)) - G(\Omega(x))\right\rVert_2 \right],
    \label{eq:loss_equivariant}
\end{equation}
where operator $\Omega$ represents an image rotation of $\ang{180}$.
Minimizing \cref{eq:loss_equivariant} ensures that pixels in $G(x)$ correspond to pixels at the same pixel coordinates in $x$ as any discrepancy would cause a mismatch of $\Omega(G(x))$ and $G(\Omega(x))$.

\subsubsection{Total loss formulation}
All components are aggregated into the \mbox{\textit{total loss}}
\begin{equation}
  \label{eq:training_objective}
  \mathcal{L} = \lambda \mathcal{L}_{R} + (1 - \lambda) \mathcal{L}_{S} + \eta \mathcal{L}_{E},
\end{equation}
with relative weightings $\lambda \in [0, 1]$ and $\eta \geq 0$ as hyperparameters.
We denote models that use Gram loss as the style loss \mbox{$\mathcal{L}_{S}$} as \mbox{\textit{\gram}}, and models that use GAN loss as the style loss \mbox{$\mathcal{L}_{S}$} as \mbox{\textit{\gan}}.
When online registration is enabled during computation of reconstruction loss $\mathcal{L}_{R}$, the corresponding models are denoted as \textit{\gramreg} and \textit{\ganreg}.
Base models \mbox{\gram} and \mbox{\gan} compute the reconstruction loss with online registration disabled.

\subsection{Model training}

For the generator $G$, we use the same 5-layer U-Net \cite{ronneberger2015} with numbers of features [32, 64, 128, 256, 512] in all experiments, and adjust the input and output channels to 3 according to our setup.
To train $G$, we use square \mpli patches of \mpx{444} size, represented by parameter maps transmittance ($I_T$), retardation ($\sin \delta$) and direction ($\varphi$).
We reformulate the \mpli parameters as triplets $(I_T, \sin \delta \cos(2 \varphi), \sin \delta \sin(2 \varphi) )$ to resolve the circular behavior of direction $\varphi$, standardize the channels and stack them to the input of generator $G$.
Due to the fully convolutional approach of the U-Net model without padding, the generated output predictions have a reduced size of \mpx{260}.
Unless specified otherwise, we use a patch size of \mpx{360} for the target \crv images, centered at the input patch position.
They are chosen to be larger than the model predictions to allow the online registration to correct translational shifts of up to \mpx{50} in any direction, while keeping the predictions fully contained within the target images.
We normalize image pixel values of the \crv staining to the range of [0, 1].

For computing style loss $\mathcal{L}_S$, we extract features from a VGG19 model \cite{simonyan2015} to compute the \textit{Gram loss}.
The VGG feature encoder network has a depth of four layers and three input channels with pre-trained weights on ImageNet \cite{deng2009}.
We multiply the style loss $\mathcal{L}_S$ by a constant factor of {10\textsuperscript{4}} to bring it to the same order of magnitude as reconstruction loss $\mathcal{L}_R$.
For the training, we use Adam  optimizer \cite{kingma2017} with $\beta_1$ = 0.9, $\beta_2$ = 0.999 and a learning rate of {10\textsuperscript{-3}}.
If not stated otherwise, we use $\eta$ = 0.1 and $\lambda$ = 0.5 as default in the training objective in \cref{eq:training_objective}.

In the case of \textit{GAN} style loss $\mathcal{L}_S$, we use a 4-layer convolutional network as discriminator with kernel size 4, stride 2, padding 1, and feature size of [32, 64, 128, 256], followed by batch normalization \cite{ioffe2015} and Leaky ReLU after each convolution.
For training, we use the Wasserstein GAN \cite{arjovsky2017a} objective and a separate Adam optimizer for the discriminator and the generator, using $\beta_1$ = 0.5, $\beta_2$ = 0.999 and a learning rate of {10\textsuperscript{-4}}.
We perform five updates for the discriminator for one update of the generator and clamp discriminator weights at 0.03 after each step.

\begin{figure}[t]
  \centering
  \includegraphics[width=0.45\textwidth]{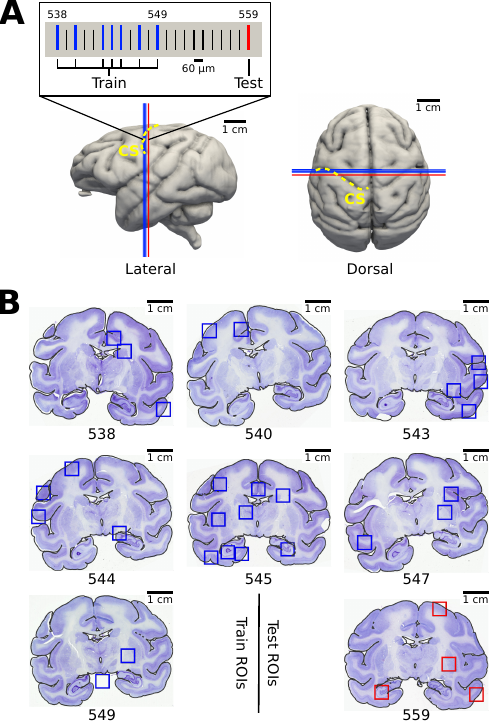}
  \caption{
    Localization of train and test data.
    \textbf{(A)} Seven sections used for training (blue stripes) and one section used for testing (red stripe) were taken at the level of the central sulcus (CS; yellow dashed lines).
    Locations are shown on top of the 3D reconstructed blockface of the brain for reference.
    Train and test data are \mmm{0.6} apart from each other.
    \textbf{(B)} Selected locations of train and test regions of interest (ROIs), which are used for training and testing the models.
    The images show ROIs from each of the train and test sections on top of globally affine registered \crv images.
    Black contour plots outline the pial surface of corresponding \mpli sections for reference.
  }
  \label{fig:traintest}
\end{figure}

\paragraph{Training data}
For model training and evaluation, we use eight coronal sections at the level of the central sulcus.
Seven sections are used for training and one section is kept for testing with a gap of $\mmm{0.6}$ between train and test sections (\cref{fig:traintest}A).
From the training sections, 27 affine-aligned ROIs are extracted (\cref{fig:traintest}B), where one ROI is held out for validation to identify possible overfitting.
For each ROI, we extract joint target \crv images and \mpli modalities at the same center location.
To maximize the diversity of the training examples, we do not pre-compute training patches but sample them randomly during the training process.
We use a batch size of 128 and draw 32,768 paired random patches per epoch evenly distributed across the training ROIs, resulting in approximately 1,260 random samples per ROI per epoch.
All training is performed for 150 epochs or until model convergence if validation loss did not decrease for at least 50 epochs.

\paragraph{Data augmentation} To enhance the robustness of our trained models we employ \mpli-specific data augmentations, which were carefully modified to keep physically plausible signal parameters \cite{oberstrass2024a}.
Specifically, we perform random rotation by angles between \mdg{-180} and \mdg{180} with mirror padding and horizontal and vertical random flipping.
In both cases, direction parameter maps $\varphi$ are corrected accordingly.
Additionally, we perform Gaussian blurring of \mpli parameter maps for random standard deviations up to $\sigma$ =1.5 and kernel sizes of 3 or 5.
We scale thickness and attenuation coefficients for \mpli parameter maps by random values between 0.5 and 2.

\paragraph{Implementation} All models were trained on the supercomputer JURECA-DC at the Jülich Supercomputing Centre \prefixcite{JSC}{thornig2021} on a single node by splitting each batch equally onto 4 NVIDIA A100 GPUs using distributed data-parallel strategy.
For data pre-processing 128 worker processes were spawned on 128 CPU cores.
For reference, training for 100 epochs took 8 hours with online registration and 4 hours without on this hardware.
The implementations are based on the Quicksetup-ai template by the HelmholtzAI Consultants Munich\footnote{https://github.com/HelmholtzAI-Consultants-Munich/Quicksetup-ai}, using the frameworks:  \textit{PyTorch} \cite{NEURIPS2019_9015}, \textit{PyTorch Lightning} \cite{jirka_borovec_2022_7447212} and \textit{Hydra} \cite{Yadan2019Hydra}.

\paragraph{Online registration}
For the online registration head, we restrict accepted solutions of \cref{eq:fourier_mse} to translation + rotation pairs that cause the registered target to have full overlap with the prediction, avoiding loss calculation over zero-padded values.
We check the translation correction for 31 rotation angles from \mdg{-7.5} to \mdg{7.5} with steps of \mdg{0.5} and take the translation + rotation pair with the best registration score.

  % Results
  \section{Experiments and Results}
\label{sec:results}

We compare a selection of performance scores to identify optimal hyperparameters of the proposed method and assess the overall potential of the best performing model.
The main hyperparameters are the choice of the online registration metric, the type of style loss $\mathcal L_S$, its relative weighting $\lambda$ to reconstruction loss $\mathcal L_R$, and whether to use the additional equivariance loss $\mathcal L_E$.
To reduce the massive computational demands by a rigorous grid search across all hyperparameters, we choose to identify a suitable choice for the online registration metric and model variants using different loss components independently before determining an optimal relative weighting $\lambda$.

\subsection{Experiment setup}

\subsubsection{Test data}

\begin{figure*}[]
  \centering
  \includegraphics[width=0.85\textwidth]{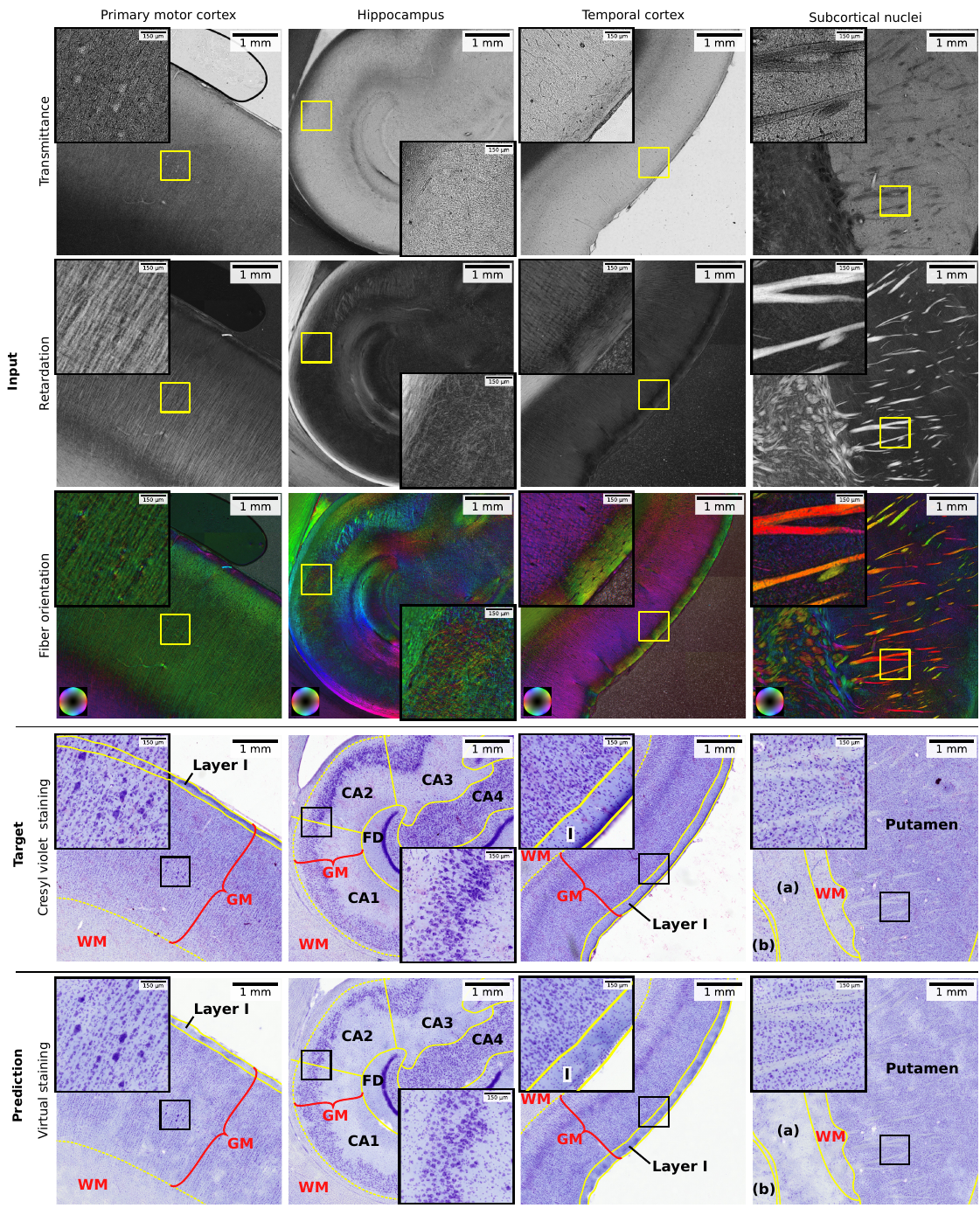}
  \caption{
    Overview of ROIs used for the evaluation, which represent distinct cellular architectures.
    They were extracted from section 559, located \mmm{0.6} apart from the training sections.
    Embedded windows show magnified details inside each ROI.
    Columns each show one of the four test ROIs taken from the anterior subdivision of the primary motor cortex (4a), the hippocampal cornu Ammonis (CA) region, temporal cortical area TE, and parts of the putamen and globus pallidus (GP; a: GP Pars interna; b: GP Pars externa) as subcortical nuclei.
    The first three rows demonstrate \mpli modalities transmittance, retardation (scaled using gamma correction for visualization) and fiber orientation in HSV color space (hue: fiber direction; saturation/brightness: retardation).
    The \mpli modalities are compared to the registered target \crv and predicted virtual staining.
  }
  \label{fig:test_images_cortex}
\end{figure*}

We manually select four ROIs for model evaluation, ensuring a diverse representation of different cytoarchitectonic characteristics from the held-out test section (\cref{fig:traintest}B).
The ROIs contain primary motor cortex 4a, temporal cortical area TE, the hippocampal cornu Ammonis (CA) region, and parts of the putamen and globus pallidus (Pars interna and Pars externa) as subcortical structures (\cref{fig:test_images_cortex}).

As the computation of image metrics requires a precise alignment of test data, we perform elastic registration of test ROIs based on landmarks and image intensity using the \textit{bUnwarpJ} \cite{arganda-carreras2006} algorithm.
We use predictions of an independently trained \gramreg model as target.
The predictions are used to manually identify 15-25 characteristic cell clusters as landmarks per ROI, which are confirmed by the location and arrangement of faint shadows of cells visible in the \mpli transmittance.
For registration, we use an image weight of 1.0, a landmark weight of 10.0, and a consistency weight of 10.0.
The strong weights for landmarks and consistency are chosen to prevent the transformation field from overly conforming to the predictions and to overcome local optima.

\subsubsection{Evaluation scores}
\label{sec:recon_quality}

We evaluate the impact of different model parameter choices on the quality of the predicted virtual staining by applying structural similarity index measure \prefixcite{SSIM}{wang2004}, mutual information (MI), and root-mean-square error (RMSE).
For each metric, we report the mean over all test ROIs.
To evaluate how well cell positions are preserved by different models, we compute F1 scores based on cell instance segmentations by a contour proposal network \mbox{\prefixcite{CPN}{upschulte2022, upschulte2023}}.
Predicted cell instances are obtained from segmentations of the virtual staining and compared to target cell instances from the corresponding \crv images.
For each ROI, predicted and target cells are matched by calculating their intersection over union (IoU), requiring a minimum IoU of 30\% for a match.
Each target cell can be a match for at most one predicted cell.
Matched instances are counted as true positives, unmatched predicted cells as false positives, and unmatched target cells as false negatives.
F1 scores are then computed from aggregated counts across all ROIs.
For the cell detection model, we fine-tuned a pre-trained CPN\footnote{ginoro\_CpnResNeXt101UNet-fbe875f1a3e5ce2c} for cell body segmentation in cell-stained microscopy images using the \textit{celldetection}\footnote{\url{https://github.com/FZJ-INM1-BDA/celldetection}} Python package.
Fine-tuning was performed on a diverse mix of manually annotated images, as well as synthetic data.

We restrict the computation of evaluation metrics to gray matter, where most of the neuronal cell bodies are located, excluding white matter and background pixels.
In test ROIs showing cortical regions, we further exclude molecular layer I as it consists of only a few individual neurons and exhibits a nonlinear deformation due to tissue shrinkage, which could not be reliably corrected in the registration.
For the ROI showing the hippocampus, we exclude the fascia dentata, as its granular layer consists of very densely packed neurons, which cannot reliably be distinguished and restrict the analysis to hippocampal CA1-CA4 regions.

\subsection{Online registration metric}
\label{sec:mse_fft_performance}

\begin{figure*}[]
  \centering
  \includegraphics[width=\textwidth]{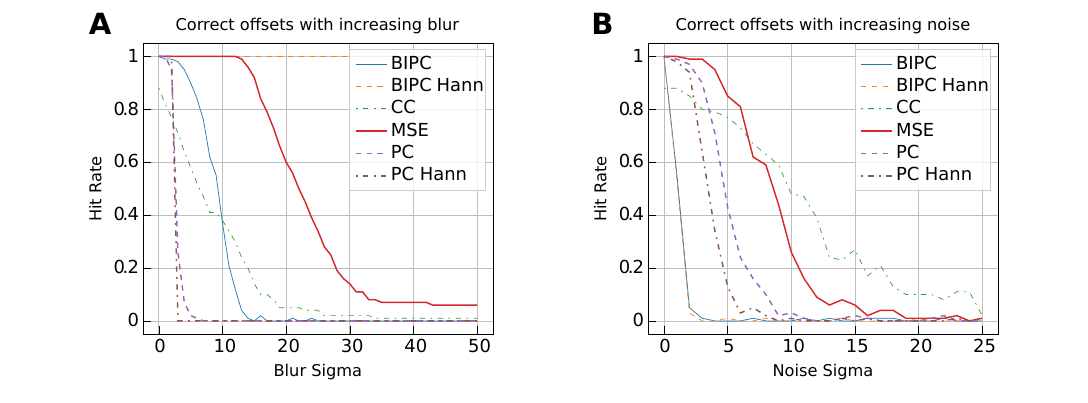}
  \caption{
    Robustness of different online registration metrics against synthetic image distortions.
    We report the proportion of correct rigid alignments for different blur (A) and noise (B) levels by registration metrics MSE, CC, PC, and BIPC.
    The registration is considered successful (a hit) if the translational displacement obtained remains within 5 pixels of the actual displacement.
  }
  \label{fig:registration_accuracy}
\end{figure*}

To compute a pixel-aligned reconstruction loss we apply the online registration head (\cref{sec:fft_reg}).
We consider correlation-based registration metrics CC and MSE with phase correlation \prefixcite{PC}{kuglin1975} and blur-invariant phase correlation \prefixcite{BIPC}{ojansivu2007}.
We apply a Hann window to PC and BIPC before their calculation to mitigate their bias towards the sharp image edges \cite{gonzalez2008digital}.

To understand how image degradation effects can  influence the success of the online registration head, we compare the robustness of the metrics against noise and blur.
A square target image with \mpx{460} size is extracted from a random location in the \crv staining.
Next, a smaller moving tile with a size of \mpx{260} is extracted from a random location within the target and distorted by gradually increasing noise or blur on the image.
We add noise from a zero-centered Gaussian distribution with standard deviation $\sigma$ increasing from 0 to 25 and Gaussian blur with increasing kernel sizes with standard deviations $\sigma$ growing from 0 to 50.
The registration head is applied with each metric to realign the degraded moving  tile with its location in the target image.
We compare the hit rate over 100 examples per degradation step.
The registration is considered successful if the determined translational displacement remains within 5 pixels of the actual displacement.

\cref{fig:registration_accuracy} shows that as blur increases, BIPC with a Hann window achieves the best performance. This result is expected, as the metric remains invariant to blur.
We also observe that CC and MSE perform best in mitigating the effects of noise, while the other metrics fail already with a minor amount of noise.
Overall, MSE provides a balanced tradeoff between blur and noise response.

\begin{table}[t]
  \centering
  \caption{
    Performance of models trained with different online registration metrics in terms of similarity of generated images with ground truth.
    We compare cross-correlation (CC), phase-correlation (PC),  blur-invariant phase-correlation(BIPC), and mean-squared error (MSE) as online registration metrics against using no online registration (-).
    Arrows indicate the direction of better performance (\mbox{$\uparrow$} higher is better, \mbox{$\downarrow$} lower is better).
    Best scores per column in bold.
   }
   \label{tab:fourier_methods_comparison}
  \begin{tabular}{cc|cccc}
    \hline \\ [-2ex]
    Method & Metric & MI $\uparrow$ & RMSE $\downarrow$ & SSIM $\uparrow$ & F1 $\uparrow$ \\
    [-2ex] \\ \hline \\ [-2ex]
    \multirow{1}{*}{\gram} & - & 0.126 & 33.6 & 0.354 & 30.2 \\
    [-2ex] \\ \cline{2-6} \\ [-2ex]
    \multirow{4}{*}{\gramreg} & CC & 0.136 & 33.0 & 0.379 & 34.4 \\
    & BIPC & 0.149 & 31.5 & 0.371 & 32.6 \\
    & PC & 0.185 & 31.9 & 0.415 & 38.0 \\
    & MSE & \textbf{0.226} & \textbf{29.8} & \textbf{0.444} & \textbf{41.3} \\ [-2ex] \\ \hline
   \end{tabular}
 \end{table}

To make an optimal choice of a registration metric for online registration, we train \gramreg models with CC, BIPC, PC, and MSE compared to a \gram model without online registration.
We use a \crv target patch size of \mpx{360} and predictions of \mpx{260}, centered within the target patch.
This allows the online registration to correct for a maximum of \mpx{50} translation in each dimension.
For \gramreg models trained with CC, BIPC and the \gram model, $\lambda = 0.1$ was chosen as it improved results compared to $\lambda = 0.5$ for PC and MSE.

\cref{tab:fourier_methods_comparison} shows a quantitative comparison of the models.
Independent of the metric, the best results are observed when using \gramreg with online registration
Among all metrics considered, the model with MSE performs best.

\subsection{Performance analysis of different model variants}
\label{seq:model_variants}

\begin{figure*}[t]
  \center
  \includegraphics[width=\textwidth]{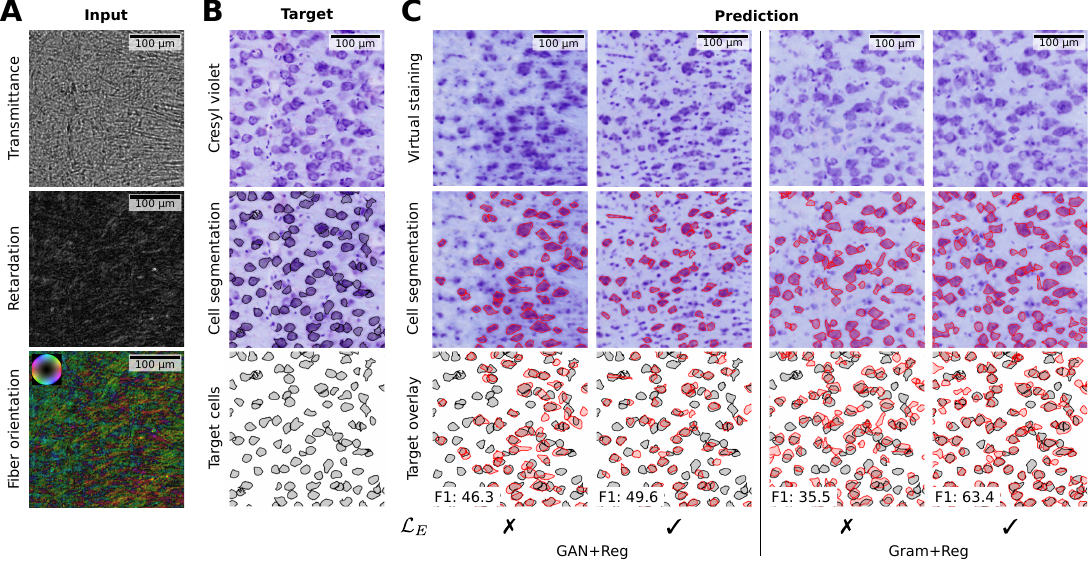}
  \caption{
    Example image illustrating how equivariance loss $\mathcal{L}_E$ improves cell instance overlap of \gramreg and \ganreg models.
    (A) \mpli input as transmittance, retardation (scaled using gamma correction) and fiber orientation map in HSV color space (hue: fiber direction; saturation/brightness: retardation).
    (B) Registered target \crv staining, a cell segmentation by a CPN model, and the identified target cell instances.
    (C) Predicted virtual staining of models trained with equivariance loss $\mathcal{L}_E$ enabled (\ding{51}) or disabled (\ding{55}).
    Predicted cell instances (red) by each model are overlaid with the target cells (gray) and F1 scores computed for cells in this specific image patch.
    Since the displacement occurs randomly, a patch from our test set and random seeds for the models were manually selected that illustrate that effect.
    The patch shows the pyramidal layer of the hippocampal CA1 region.
  }
  \label{fig:equivariant_cpn}
\end{figure*}

To justify including equivariance loss $\mathcal{L}_E$ (\cref{eq:loss_equivariant}) in the total loss, \cref{fig:equivariant_cpn} illustrates the effect of training different models with this loss component enabled and disabled.
With $\mathcal{L}_E$ disabled ($\eta$ = 0), cell instances in the model predictions exhibit a relative displacement.
Enabling $\mathcal{L}_E$ ($\eta$ = 0.1) improves the overlap of cell instances with the original \crv staining.

\begin{table*}[]
  \centering
  \caption{
    Effect of the proposed online registration on models trained with different choices of style loss $\mathcal L_S$.
    Each variant is trained with online registration head (Reg.) and equivariance loss ($\mathcal{L}_E$) enabled or disabled.
    MSE metric is used for online registration and $\eta$ = 0.1 for weighting of $\mathcal{L}_E$.
    The deviation is reported as standard error over four independent trainings with different random seeds.
    Arrows indicate the direction of better performance (\mbox{$\uparrow$} higher is better, \mbox{$\downarrow$} lower is better).
    Best scores per column in bold.
 }
 \label{tab:eq_effect}
  \begin{tabular}{ccc|cccc}
    \hline \\ [-2ex]
    $\mathcal L_S$ & Reg. & $\mathcal{L}_E$ & MI $\uparrow$ & RMSE $\downarrow$ & SSIM $\uparrow$ & F1 $\uparrow$ \\ [-2ex] \\ \hline \\[-2ex]
    \multirow{4}{*}{\gram} & \multirow{2}{*}{\ding{51}} & \ding{51} & \textbf{0.224} $\pm$ 0.002 & \textbf{29.6} $\pm$ 0.4 & \textbf{0.445} $\pm$ 0.002 & \textbf{41.0} $\pm$ 0.4 \\
    & & \ding{55} & 0.211 $\pm$ 0.004 & 30.2 $\pm$ 0.4 & 0.432 $\pm$ 0.004 & 35.2 $\pm$ 1.2 \\ 
    [-2ex] \\ \cline{2-7} \\ [-2ex]
    & \multirow{2}{*}{\ding{55}} & \ding{51} & 0.111 $\pm$ 0.012 & 33.6 $\pm$ 0.8 & 0.314 $\pm$ 0.023 & 22.6 $\pm$ 4.8 \\
    & & \ding{55} & 0.107 $\pm$ 0.007 & 34.2 $\pm$ 0.3 & 0.304 $\pm$ 0.015 & 22.5 $\pm$ 3.1 \\
    [-2ex] \\ \hline \hline \\ [-2ex]
    \multirow{4}{*}{\gan} & \multirow{2}{*}{\ding{51}} & \ding{51} & 0.160 $\pm$ 0.008 & 30.5 $\pm$ 0.5 & 0.389 $\pm$ 0.012  & 34.7 $\pm$ 1.6 \\
    & & \ding{55} & 0.126 $\pm$ 0.008 & 33.0 $\pm$ 0.4 & 0.347 $\pm$ 0.010 & 28.0 $\pm$ 1.3 \\
    [-2ex] \\ \cline{2-7} \\ [-2ex]
    & \multirow{2}{*}{\ding{55}} & \ding{51} &  0.085 $\pm$ 0.008 & 38.1 $\pm$ 0.5 & 0.249 $\pm$ 0.001 & 12.9 $\pm$ 0.3 \\
    & & \ding{55} & 0.096  $\pm$ 0.008 & 38.8 $\pm$ 1.7 & 0.253 $\pm$ 0.005 & 13.9 $\pm$ 1.5 \\
    [-2ex] \\ \hline
    \end{tabular}
\end{table*}

To quantify the impact of the loss component $\mathcal{L}_E$ on the predictions, we compare the models with $\mathcal{L}_E$ enabled and disabled in \cref{tab:eq_effect}.
We consider \gram and \gramreg, as well as \gan and \ganreg as model variants.
All models use $\lambda = 0.5$ expect for the \gram model without online registration, where $\lambda = 0.1$ showed better performance.
Since displacements of cell instances in the model predictions occur randomly with varying strength, we report mean and standard errors across four independent trainings.
Models trained with $\mathcal{L}_E$ enabled show the highest correspondence with the target staining.
Without online registration, the advantage of including $\mathcal{L}_E$ becomes negligible.
Furthermore, results in  \cref{tab:eq_effect} demonstrate that \gram models outperform their corresponding \gan models in all evaluation metrics.
Regardless of the variant used for style loss $\mathcal{L}_S$, all models benefit from the online registration.
We present averaged scores across all test ROIs.
Individual scores per ROI, are provided in \mbox{\cref{tab:all_rois}} in the appendix as additional information.

\begin{figure*}[]
  \center
  \includegraphics[width=\textwidth]{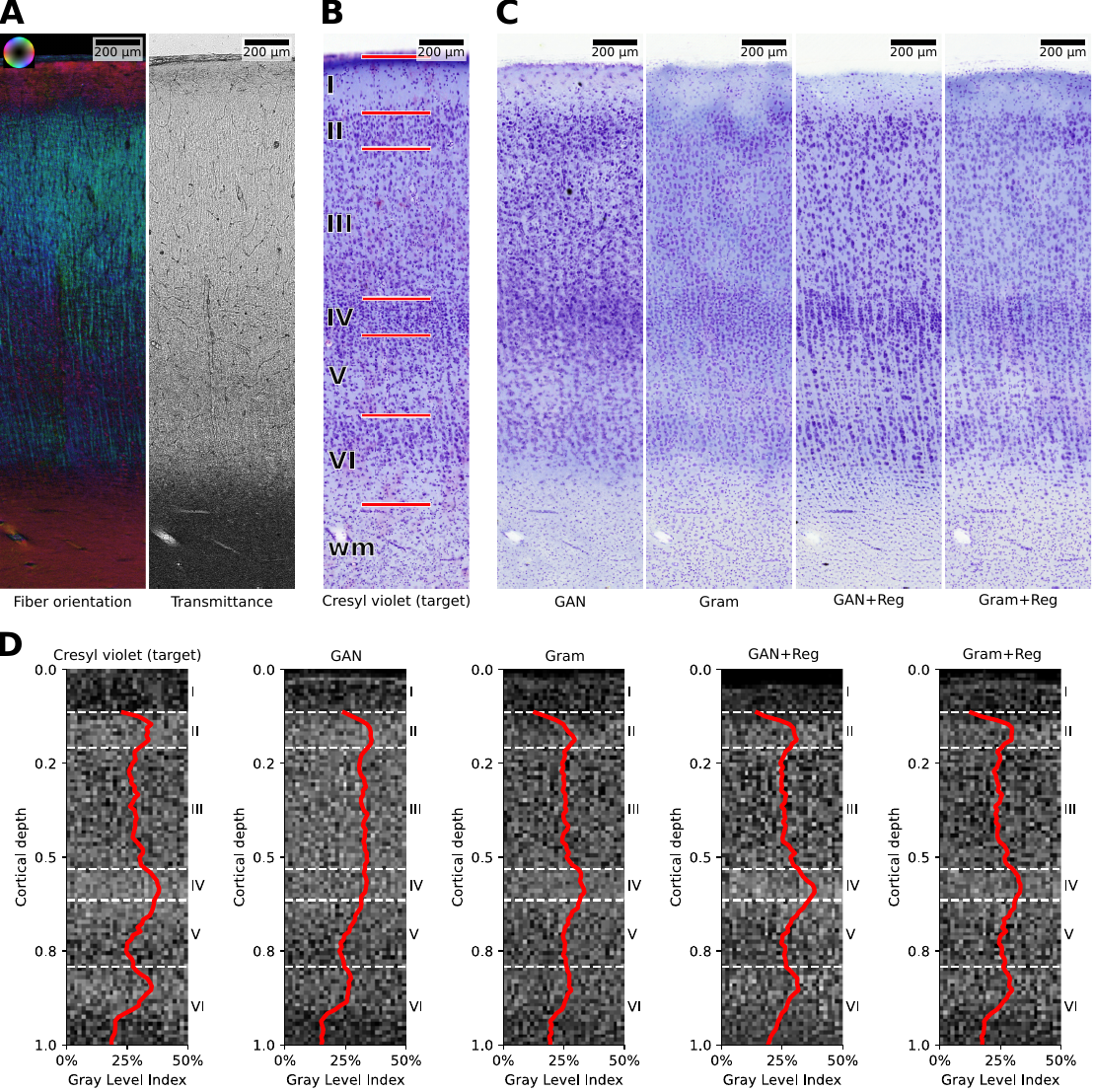}
  \caption{
    Comparison of virtual stainings for a patch of temporal area TE.
    (A) \mpli input visualized as fiber orientation in HSV color space (hue: fiber direction; saturation/brightness: retardation) and transmittance.
    (B) Registered target \crv staining for reference with annotations of cortical layers I-VI and white matter (wm).
    (C) Predicted virtual stainings of \gan and \gram models and extended variants using MSE online registration (\ganreg, \gramreg).
    (D) GLI images ranging from the pial surface to the cortex/white matter transition and average profiles (red lines) for layers II - VI.
  }
  \label{fig:gram_vs_gan}
\end{figure*}

A comparison of model predictions is shown in \cref{fig:gram_vs_gan}C for the whole cortical depth of an Isocortex sample from temporal cortical area TE.
\ganreg and \gramreg provide the most realistic-looking reconstructions in terms of relative size and shape of cell bodies, along with the differences between layers concerning cell packing density.
While \gramreg seems to be influenced by technically-related inhomogeneities in staining intensity, \ganreg produces a clearer contrast between stained cells and surrounding tissue.
\gramreg detects the transition from the cortex to white matter better than \ganreg.
Both methods introduce an artificial arrangement of cells into cortical columns not present in the original \crv staining.

To further analyze the ability of the methods to reconstruct the laminar cell organization, we compute grey level index (GLI) values \cite{zilles1978}.
GLI values provide an established proxy for volume density of stained cell bodies in gray matter regions and are used to characterize the laminar architecture of cortical areas \cite{schleicher2000}.
To compute GLI values, adaptive thresholding is applied to create a mask of pixels occupied by cell bodies for each image.
The masks are subsequently down scaled by a factor of 16 to a spatial resolution of \mmu{20.8}, with each value representing the fraction of segmented pixels.
Resulting GLI images, cropped to the area between pial surface and the cortex/white matter transition, are displayed for the \crv target and the predictions in \cref{fig:gram_vs_gan}D.

For each of the GLI images, 31 intensity profiles vertically oriented to the cortical layers are extracted to represent the columnar distribution of GLI values.
The profiles are restricted to layers II - VI since staining inhomogeneities make GLI values for layer I unreliable.
To obtain one representative profile for each image, the profiles are averaged and smoothed using a mean filter with kernel size 3 to improve the signal-to-noise ratio.
The average profiles are shown on top of each GLI image in \cref{fig:gram_vs_gan}D.

The average GLI profile for the \gan model shows no clear distinction between cortical layers II-IV and misses a peak for layer IV.
The profile by \gram replicates peaks for layers II and IV but does not provide a clear representation of the cortical layers.
\ganreg and \gramreg both replicate peaks for the higher cell densities in layers II, IV and VI.
\gramreg shows the clearest distinction between layers and even replicates slightly nuanced peaks of the profile within layers.

\subsection{Effect of loss weighting parameters}
\label{seq:style_balancing}

\begin{table}[]
  \centering
  \caption{
    Ablation study on weighting parameter \mbox{$\eta$} of equivariance loss \mbox{$\mathcal{L}_E$} for a \mbox{\gramreg} model.
    Performance is robust across a wide range of \mbox{$\eta > 0$} values.
    Only when the loss term is completely removed \mbox{($\eta = 0$)}, we observe a measurable degradation in performance.
    The deviation is reported as standard error over four independent trainings with different random seeds.
    Arrows indicate the direction of better performance (\mbox{$\uparrow$} higher is better, \mbox{$\downarrow$} lower is better).
    Best scores per column in bold.
 }
 \label{tab:ablation_eta}
  \begin{tabular}{cc|cccc}
    \hline \\ [-2ex]
    Method & $\eta$ & MI $\uparrow$ & RMSE $\downarrow$ & SSIM $\uparrow$ & F1 $\uparrow$ \\ [-2ex] \\ \hline \\[-2ex]
    \multirow{5}{*}{\gramreg}
      & 0.0  & 0.211 $\pm$ 0.004 & 30.2 $\pm$ 0.4 & 0.432 $\pm$ 0.004 & 35.2 $\pm$ 1.2 \\
      & 0.01 & \textbf{0.225} $\pm$ 0.002 & 29.5 $\pm$ 0.3 & \textbf{0.446} $\pm$ 0.001 & \textbf{41.6} $\pm$ 0.2 \\
      & 0.1  & 0.224 $\pm$ 0.003 & 29.6 $\pm$ 0.6 & 0.445 $\pm$ 0.003 & 41.0 $\pm$ 0.4 \\
      & 1.0  & 0.220 $\pm$ 0.001 & \textbf{29.4} $\pm$ 0.1 & 0.445 $\pm$ 0.001 & 41.4 $\pm$ 0.2 \\
      & 10.0 & 0.224 $\pm$ 0.002 & \textbf{29.4} $\pm$ 0.3 & \textbf{0.446} $\pm$ 0.001 & 41.2 $\pm$ 0.2 \\
    [-2ex] \\ \hline
  \end{tabular}
\end{table}

To incorporate equivariance loss $\mathcal{L}_E$ into previous models, we set its weighting parameter to $\eta = 0.1$.
To validate this choice, we perform training of \mbox{\gramreg} models with $\eta$ values ranging from 0 to 10.
Results in \mbox{\cref{tab:ablation_eta}} indicate that performance is robust for $\eta > 0$.
Only when the equivariance loss is entirely removed from training (setting $\eta = 0$), we observe a measurable negative effect on the results.
We report mean values and standard errors over four independent trainings to show the significance of this effect.

\begin{table}[]
  \centering
  \caption{
    Different values for weighting parameter $\lambda$ show that a balance between style loss $\mathcal{L}_S$ and reconstruction loss $\mathcal{L}_R$ is required in the training of a \gramreg model.
    Setting $\lambda = 0$ means only $\mathcal{L}_S$ is used. Setting $\lambda = 1$ means only $\mathcal{L}_R$ is used.
    Arrows indicate the direction of better performance (\mbox{$\uparrow$} higher is better, \mbox{$\downarrow$} lower is better).
    Best scores per column in bold.
 }
  \label{tab:style_weights}
  \begin{tabular}{cc|cccc}
    \hline \\[-2ex]
    Method & $\lambda$ & MI $\uparrow$ & RMSE $\downarrow$ & SSIM $\uparrow$ & F1 $\uparrow$ \\
    [-2ex] \\ \hline \\[-2ex]
    \multirow{7}{*}{\gramreg} & 0 & 0.124  & 33.8 & 0.340 & 28.6 \\
    & 0.03 & 0.162 & 31.9 &  0.394 & 36.1 \\
    & 0.25 & 0.210 & 30.7 & 0.436 & 39.8 \\
    & 0.5 & 0.226 & \textbf{29.8} & 0.444 & 41.3 \\
    & 0.75 & \textbf{0.238} & \textbf{29.8} & \textbf{0.448} & \textbf{41.4} \\
    & 0.97 & 0.230 &  30.7 & 0.446 & 40.2 \\
    & 1.0 & 0.090 & 38.0 & 0.347 & 0 \\
    [-2ex] \\ \hline
  \end{tabular}
 \end{table}

 \begin{figure}[]
  \center
  \includegraphics[width=\textwidth]{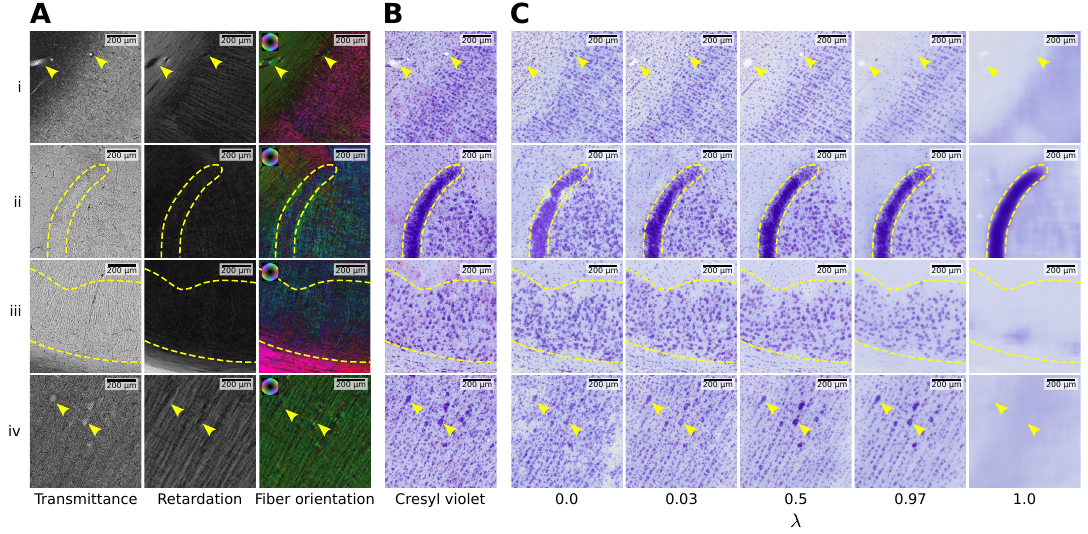}
  \caption{
    Predicted virtual \crv stainings for different weightings between reconstruction  and style loss components.
    (A) \mpli inputs as transmittance, retardation and fiber orientation in HSV color space (hue: fiber direction; saturation/brightness: retardation).
    The retardation has been scaled using gamma correction for visualization.
    (B) Registered target \crv staining.
    (C) Predicted virtual stainings of models trained with different weightings $\lambda$ between reconstruction loss $\mathcal{L}_R$ and style loss $\mathcal{L}_S$.
    Setting $\lambda = 0$ focuses on $\mathcal{L}_S$ and setting $\lambda = 1$ on $\mathcal{L}_R$ exclusively.
    Gram loss is used for $\mathcal{L}_S$.
    Within each row, yellow markers indicate the same structures across all columns.
    (i) Temporal area TE and underlying white matter with arrowheads highlighting blood vessels within white matter and cortical layer VI.
    (ii) Dentate gyrus of the hippocampus with a dashed line delineating the proximal end of the granular layer of the fascia dentata (FD).
    (iii) Pyramidal layer of the cornu Ammonis (CA1) region of the hippocampus highlighted by dashed lines.
    (iv) Layer V of the primary motor cortex with arrowheads highlighting two Betz cells.
  }
  \label{fig:compare_style_loss_weight}
\end{figure}

To identify an optimal weighting $\lambda$ of the style loss in \cref{eq:training_objective}, we train \gramreg models with different values for $\lambda$ and MSE online registration metric.
A quantitative evaluation of the trained models in \cref{tab:style_weights} shows that larger values for $\lambda$ up to $\lambda$ = 0.75 achieve the best evaluation scores.
This indicates that the reconstruction loss should be given a stronger weight but should not be used exclusively ($\lambda = 1$).
% The highest overlap of predicted cell instances measured by the F1 score is obtained for a balanced weighting with $\lambda$ = 0.5.

A visual comparison of predictions of \gramreg models trained with different choices for $\lambda$ is shown in \cref{fig:compare_style_loss_weight} for selected crops from the test data.
By using $\mathcal{L_R}$ only (i.e. $\lambda = 1$), the model is unable to reconstruct details from the \crv staining.
It is only able to locate strongly pronounced structures such as the granular layer delineated in (ii).
Independent of the style loss weighting, the models do not resolve individual cell instances present within the granular layer.
Instead, they capture the overall cell density, represented by a continuous dark purple color.
By increasing emphasis on $\mathcal{L}_S$ (i.e. with decreasing values for $\lambda$), the generator is able to reconstruct details such as neuronal cell bodies in less dense regions in (i)-(iv) and blood vessels in (i).
It is also able to generate Glial cells to match the appearance of white matter in (i) and (iii).
However, generators trained with an overweighting of style loss also tend to miss some strongly pronounced structures in the predictions, such as the blood vessels shown in (i) or the Betz cells in (iv).

\subsection{Influencing factors on model predictions}
\label{seq:reliability}

To assess the reliability of our method under varying biological and imaging conditions, we analyze in how far the reconstruction of cell instances depends on cell size and local strength of birefringence.
We also examine how different focus levels of the LMP-1 impair the quality of the virtual staining.

\begin{figure}[]
  \center
  \includegraphics[width=\textwidth]{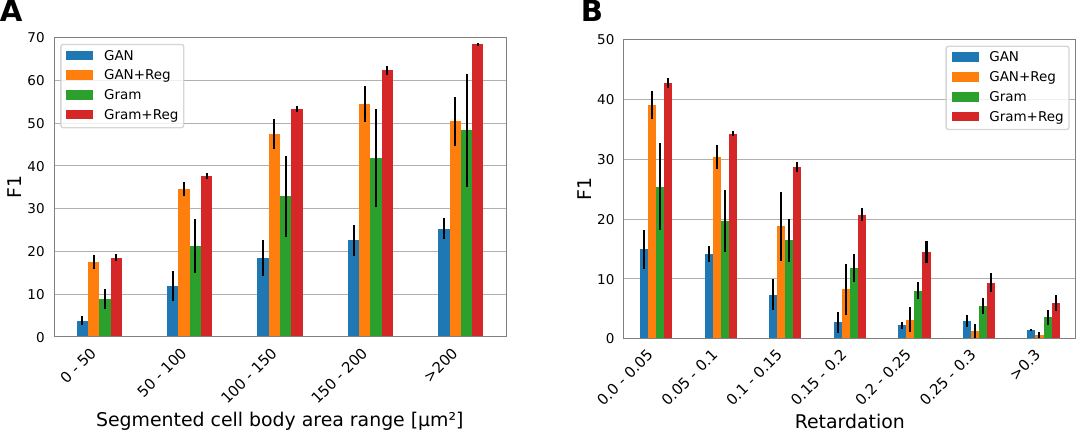}
  \caption{
    Reliability of predicted cell instances increases with their size and decreases with a stronger retardation.
    We report F1 scores between detected cell instances by a CPN model in the predicted and target stainings.
    F1 scores are computed for intervals of (A) the segmented in-plane cell body area and (B) smoothed retardation maps.
    Retardation values are sampled at each cell location.
    Each bar represents the average over four independently trained models.
    Error bars show standard deviation.
  }
  \label{fig:f1_by_factors}
\end{figure}

Larger cells are expected to be more pronounced in \mpli parameter maps compared to smaller cells that may be overshadowed by other tissue components such as nerve fibers.
To quantify the reliability of predicted cells in the virtual staining in relation to their size, we compute F1 scores for multiple bins of 2D in-plane cell sizes.
Obtaining F1 scores for each bin requires computation of true and false positives, as well as false negatives through matching of predicted and target cells with a minimum IoU threshold of 0.3.
We modify the computation of true and false positives by matching only predicted cells within that range with all cells in the target image.
To count false negatives, target cells within that range are matched with all cells in the prediction.
Results in \mbox{\cref{fig:f1_by_factors}A} show that reconstruction of smaller cells < \msmu{50} has much lower F1 scores, below 20.0, throughout all methods.
With increasing 2D cell size, they can be identified more accurately.
The \gramreg model shows highest F1 scores across all cell sizes and smallest variation between four independently trained models, measured as standard deviation.

Birefringent tissue, such as myelinated nerve fibers, can obscure signals from other components in \mpli.
To examine whether this effect impacts the prediction of cell instances in the virtual staining, we use retardation maps as a measure of birefringence strength.
The maps are smoothed with a \mbox{\mpx{10}} square median kernel to reduce local variance and obtain values representative of the surrounding tissue area.
For each cell instance segmented by the CPN model in the virtual staining, we sample a retardation value at its center location.
Values are grouped into intervals from 0.0 to 0.3 in steps of 0.05.
Since we focus the analysis on gray matter, larger retardation values do not occur.
As shown in \cref{fig:f1_by_factors}B, the ability of all models to reconstruct cell instances decreases significantly for cells dominated by stronger retardation signals.
For \gramreg this effect is minimal, performing overall best.

\begin{figure*}[]
  \centering
  \includegraphics[width=\textwidth]{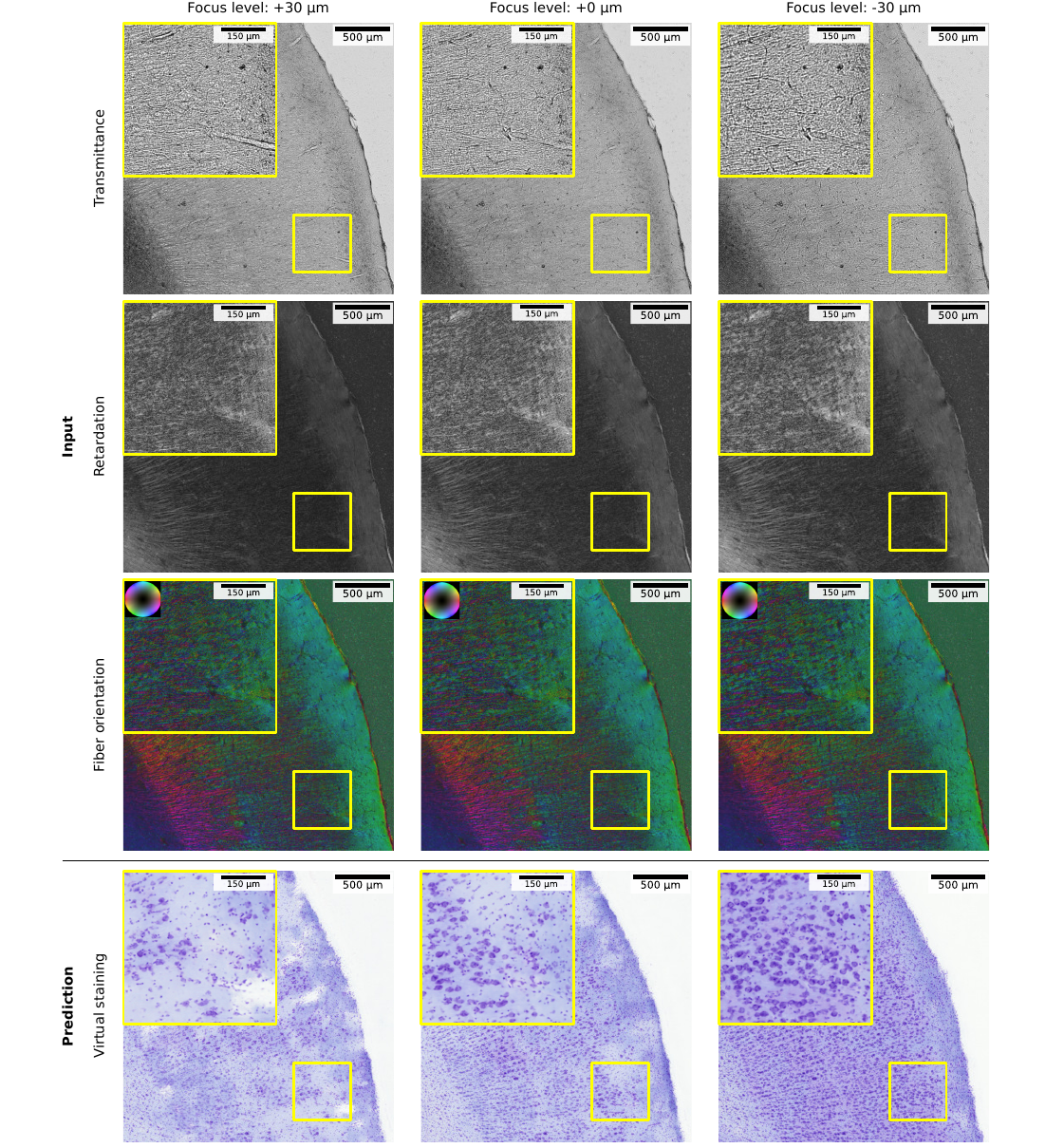}
  \caption{
    Virtual staining quality varies with the focus level of the LMP-1 microscope, which influences the captured diffraction patterns.
    Example images show the vervet entorhinal cortex with zoom-ins to the architecture of cortical layers II-III.
    \mpli inputs are represented as transmittance, retardation (scaled with gamma correction) and fiber orientation in HSV color space (hue: fiber direction; saturation/brightness: retardation).
    The same \gramreg model is used to predict virtual stainings from a focus level adjusted to the tissue center (+\mmu{0}) and \mmu{30} above and below.
  }
  \label{fig:focus_levels}
\end{figure*}

We observe that the virtual staining intensities are not always homogenous within generated images.
Such effects can especially be observed as cloud-like patterns in \cref{fig:test_images_cortex} for the predicted virtual staining of ROIs showing the primary motor cortex and subcortical nuclei.
These variations coincide with variations in the diffraction pattern of the associated transmittance maps, which in turn can be influenced by the focus level of the LMP-1 microscope.
To better quantify this relationship, we apply a \gramreg model to predict virtual stainings from an example image showing the vervet entorhinal cortex, captured at three focus levels: centered on the tissue (\mbox{+\mmu{0}}), \mbox{-\mmu{30}} and \mbox{+\mmu{30}}.
Results are summarized in \cref{fig:focus_levels}.
We observe the highest consistency in staining intensity at the lowest focus level of {(\mbox{-\mmu{30}})}.
Higher focus levels introduce an increasing amount of staining artifacts.

  % Discussions
  \section{Discussion}
\label{sec:discussions}

Our ablation experiments in \cref{sec:mse_fft_performance}, \cref{seq:model_variants} and \cref{seq:style_balancing} support the design decisions of the proposed virtual staining method.
In \cref{seq:reliability} we investigated several influencing factors on the predictions.
In the following, we want to discuss the design decisions of the proposed virtual staining method and assess its reliability for downstream analysis.

We introduced an online registration head capable of approximating smooth nonlinear deformations, given sufficiently small patch sizes for training.
Its accuracy is inherently influenced by the choice of the registration metric.
We identified MSE in the Fourier domain through multiple applications of cross-correlation as superior to other typically used metrics in Fourier-based image registration, including PC.
This is a notable observation, as PC is a popular choice due to its robustness to intensity variations and frequency-dependent noise \cite{tong2019}. 
However, during early training stages, model predictions are often blurred. The network tends to learn coarse, low-frequency structures first. Additionally, random weight initialization introduces noise in the first steps. MSE appears more resilient to such blur and frequency-independent noise, leading to more accurate predictions (\cref{fig:registration_accuracy}).
This behavior can be explained by PC’s reliance on strong high-frequency spectral peaks, which are diminished by blur \cite{pedone2013}. MSE, in contrast, maintains sensitivity across the frequency spectrum and can better handle smooth, low-detail inputs that dominate during initial training.

The applied pixel-wise reconstruction loss assumes a precise alignment to provide an informative feedback signal. 
However, the online registration head will typically only succeed if a good contrast between structures is being predicted.
In other words, there is a cross-dependency of accurate pixel reconstructions and producing a high contrast over training iterations, which may result in a conflict during training.
This can be seen in \cref{fig:compare_style_loss_weight}, where a \gramreg model focussing on pixel reconstruction only made heavily blurred predictions that do not provide structural details for cell-precise registration.
Adding a style loss component leads to more pronounced contours while simultaneously achieving higher pixel reconstruction accuracy (\cref{tab:style_weights}) through online registration.
This underpins the importance of the additional style loss, which does not depend on perfect alignment and guides the training to produce structures.
Still, some manual pre-alignment was required to keep the registration error of training data within \mbox{\mpx{70}} \mbox{(\mmu{91})} and ensure sufficient structural overlap between training patches.
This limited the amount of available training data, as larger registration errors, such as those from the initial affine alignment of whole sections, were too large for the online registration to succeed.

Apparently, the choice for Gram or GAN as  style loss should be considered in the light of the application.
Training with Gram loss led to more accurate cell localization and pixel values (\cref{tab:eq_effect}).
However, it was affected by staining intensity inhomogeneities, resulting in blurry cells (\cref{fig:gram_vs_gan}). 
GAN loss, on the other hand, produced higher contrast in cells but with lower accuracy.
This could be a consequence of the Gram loss referring to the specific target image, while the GAN loss refers to the average scoring of the discriminator across the distribution, less than the precision with respect to an individual image.

The choice of style loss function also impacts the stability and complexity of the training process.
When using GAN loss, the training process has to be carefully monitored.
If training collapses, restarting is often necessary.
It requires careful balancing of discriminator and generator capacities and tuning of style and reconstruction loss weighting.
Training with \gram loss, on the other hand, is generally more convenient to train.
It requires no balancing of model capacities, is robust to loss balancing, and does not collapse during training.

We expect that the expressiveness of the Gram matrix representation could be enhanced by replacing the VGG encoder, pre-trained on ImageNet, with a domain-specific model, which is powerful on microscopy data.
This is technically motivated by the observation that higher-level features in ImageNet-pretrained networks, such as VGG, are optimized for object semantics in natural images (e.g., faces, animals, vehicles), which are irrelevant in the context of fine-grained texture characteristic of microscopy images \cite{stuckner2022}.
In contrast, microscopy-specific encoders are better suited to capture such low- to mid-level textural features and have been shown to improve clustering \cite{oberstrass2024a}, segmentation \cite{stuckner2022}, and the evaluation of generative models \cite{kropp2024} over those trained on ImageNet.
Therefore, computation of Gram matrices using domain-specific encoders \cite{spitzer2017, schiffer2021, schiffer2021a} or emerging foundation models for histology \cite{tran2024} might further enhance representation of cytoarchitectural characteristics.

Throughout all models, we observed occasional staining inhomogeneities in the predicted virtual stainings.
An example of local-scale variation is the inhomogeneity observed in \mbox{\cref{fig:gram_vs_gan}}, which may be attributed to the tiling strategy used during model inference to transfer whole sections.
Since the model lacks a global view of the section, it may reproduce variations in the training data at the local patch level.
At a larger scale, this effect was pronounced in the form of stripe patterns, for example in the primary motor cortex ROI (\cref{fig:test_images_cortex}).
These patterns typically coincide with inhomogeneities in the \mpli transmittance maps.
During the mounting process of the \mmu{60} thick sections in 20\% glycerol solution on the object slide the gray matter is more prone to swelling than the more compact white matter, leading to tissue expansion, mechanical stress and tape flutter especially in the cortex due to its higher flexibility.
The free-floating tissue between object slide and coverslip shapes waves of up to \mmu{30} in amplitude.
These waves affect the focus level and the light incidence on the tissue, hence the effective birefringence, modifying the polarization state of the transmitted light.
This caused the observed stripe patterns in the transmittance maps and very likely contributed to the staining inhomogeneities.

Cell body membranes and blood vessel walls pose hard edges in the cortex, resulting in diffraction patterns even brighter than the background field.
Varying of the focus level of the LMP-1 microscope allows optical scanning of the Poisson spots induced by diffraction.
Cell bodies are highlighted with higher contrast by shifting and lowering the focal plane to the top and the bottom of the tissue by approximately \mbox{$\pm$\mmu{30}} (\cref{fig:focus_levels}).
The ability to predict cells from \mpli measurements was shown to be sensitive to variations in the focus level.
An effective countermeasure against waves in the tissue, to maintain a constant level across the entire section, is to weight the open coverslips after embedding overnight.
In addition, the acquisition of multiple levels for each section could potentially provide missing information.
For the given dataset, however, these potential solutions were not applicable, and measurements could not be easily repeated.
Since shifting the focus toward the tissue surface degrades optical resolution, a center focus level is used by default in \mpli for an optimal representation of nerve fibers.
The acquisition of multiple focus levels of \mbox{$\pm$\mmu{30}} to increase the visibility of cell bodies would significantly increase the measurement effort.
Nevertheless, it will be an important consideration for future investigations to include multiple focus levels in each measurement to improve cell extraction.

It must be noted that shrinking and swelling of the tissue also affects the measured areas of individual 2D segmented cell bodies.
Tissue deformation in histology is complex, involving anisotropic effects and differences between cellular and extracellular structures \cite{dorph-petersen2001,west2013}, which cannot be fully captured by the global shrinkage factor estimated in \cref{sec:tissue_shrinkage_correction}.
At the same time, there is only sparse data available that quantifies such effects, and the precise conditions of histological processing differ between each other, and it can be assumed that such differences influence the specific shrinkage values.
In paraffin-embedded human sections, a mean difference of 9\% between gray and white matter shrinkage has been reported \cite{kretschmann1982}.
As frozen sections generally show less in-plane deformation than paraffin sections \citep{west2013}, we treat this value as an upper bound for our acqusition.
Since gray matter contains relatively more cell bodies than white matter, similar shrinkage of gray and white matter suggests that shrinkage of cell bodies and neuropil should also not differ significantly \cite{amunts1999}.
By performing non-linear co-registration of \crv to \mpli and applying a subsequent global correction of cell sizes to account for swelling effects from MRI to \mpli, we assume that major shrinkage effects are compensated.
However, residual local variations or relative slice-to-slice differences may persist and will require a more systematic approach in future work.

While most of the larger cells (>\msmu{100} in-plane size) could be localized in \mpli parameter maps, smaller cells were often missed or misplaced in the predictions.
This is expected as smaller cell structures are often dominated by a stronger signal of intersecting nerve fibers or overlay each other due to the high section thickness of \mmu{60} required for \mpli.
In our experiments, we observed that stronger birefringence, corresponding to higher retardation values, resulted in reduced capability of models to reconstruct cell instances.
Since birefringence increases with myelin density and homogeneity, predictions in regions with densely packed myelinated nerve fibers, such as white matter, are thus less reliably captured by the virtual stainings.
In white matter, detecting cell bodies may even be infeasible, as the predominant oligodendrocytes form the myelin sheath and are therefore indistinguishable from it.
In gray matter, this limitation may explain the stronger arrangement of cells into cortical columns in the predictions compared to the target images in \cref{fig:gram_vs_gan}, as the strong signal of myelinated radial fibers can obscure underlying cell instances.
The presence or absence of this arrangement is one of the criteria used to identify cortical areas, which might impair downstream interpretation of the predicted architecture.

As shown by the differences in F1 values in \cref{tab:all_rois}, cells were reconstructed more accurately in isocortical areas (motor cortex and temporal cortex) than in the hippocampus or the subcortical nuclei (putamen and globus pallidus).
The lower F1 in the putamen compared to the motor and temporal areas can be attributed to its neuronal composition and the fundamentally different spatial arrangement of cell bodies in cortical versus subcortical regions, as well as their relation to fiber bundles.
The putamen is composed of medium-sized spiny neurons and relatively small interneurons arranged around and between bundles of myelinated fibers, which course through the putamen and give rise to its characteristic striated appearance \cite{heilbronner2025}.
In this region, two challenges reduce predictability: small interneurons are difficult to detect with our approach, and fiber bundles may intersect or obscure cell bodies, making them harder to reconstruct.
Moreover, the cell bodies of spiny neurons can resemble fiber bundles oriented perpendicular to the section plane, appearing as dark patches in the transmittance image.
The globus pallidus contains relatively large but sparsely distributed neurons, which are embedded in a dense matrix of multidirectionally oriented myelinated fibers \cite{heilbronner2025}.
Here, predictability is reduced by the number and complex spatial organization of intersecting myelinated fiber bundles and blood vessels.
The low F1 value in patches extracted from the CA1-CA3 regions of the hippocampus is more difficult to explain, though it may result from the fact that they cover both the pyramidal layer, with its numerous and relatively large pyramidal cells, and the radiatum layer, which contains sparsely packed and relatively small interneurons \cite{zhao2025}, which are difficult to detect with our method.

Our \gramreg model produced larger cells at expected positions and generated images with a plausible appearance, making virtual stainings useful for cross-modality registration - at present the most relevant application.
They also allowed the application of cytoarchitectonic tools such as cell segmentation or computation of GLI profiles for cortical layer characterization on \mpli images.
As a future perspective, the ability to detect cell bodies in \mpli may enhance the computation of 3D fiber orientations within gray matter, which is a prerequisite for fiber tractography.
This could be achieved by improving the estimation of myelination, for example through the identification of voxels dominated by cell bodies, and potentially also support the localization of axon terminals.
However, virtual stainings occasionally contained artifacts, including staining inhomogeneities, omission of smaller cells, or the introduction of implausible cytoarchitectonic features.
These issues may reflect biases in the training data, model-related artifacts, or missing cellular signatures in the \mpli parameter maps.  
The virtual staining allowed the identification by a neuroanatomist (N.P.-G.) of the borders between the CA1, CA2, and CA3 regions of the hippocampus, the border between the primary motor and primary somatosensory areas, or the border between the retrosplenial cortex and cingulate area 23.
However, it was not possible to identify the border between the core and lateral belt auditory areas, because it is characterized by differences in the packing density of small pyramids \cite{hackett2001}, which currently cannot be reliably predicted.  
Therefore, we do not yet consider the method sufficiently robust for reliable cytoarchitectonic brain area mapping.
Nevertheless, we are convinced that expanding the number of training sections, brain regions, and focus levels across brains and species can improve model performance and generalizability.
While such expansion would require re-training, each additional sample would teach the model novel architectural patterns, improving its robustness across domains.

  % Conclusions
  \section{Conclusions}
\label{sec:conclusions}

Motivated by previous observations that larger cells are encoded in \mpli parameter maps alongside fiber orientations \cite{zeineh2017}, we introduced a deep learning model for transforming \mpli maps into virtual \crv cell body stainings.
This approach enables joint visualization of fiber tracts and cell bodies in the same tissue.
Compared to real post-staining, the model may offer a scalable alternative that avoids manual labor.

A central contribution of our approach is the integration of an online registration head during training.
This component eliminates the need for explicit, pixel-accurate multimodal registration, which is commonly required in virtual staining pipelines \cite{rivenson2020,dehaan2021,yang2022}.
It is a simple but highly efficient add-on that can be combined with various loss formulations, leveraging model-estimated landmarks, to continuously refine the alignment over time.

The developed method enables localization of most of the larger cell bodies in gray matter (\mbox{>\msmu{100}} in-plane size) from \mpli and a successful adaptation to the appearance of real \crv stainings.
As such, it expands the usability of \mpli in large-scale data settings, allowing virtual staining at scale.
While such synthetic data cannot and should not replace real histological measurements, it offers promising opportunities in downstream analysis.
Although we see the present method not yet robust enough for cytoarchitectonic mapping, potential applications include cross-modal image registration to align real \crv and \mpli, performing cell segmentation in \mbox{\mpli} images, or missing data imputation in serial section stacks.
Especially in interleaved modalities, this could enable the reconstruction of complete datasets.
Of course, such applications require careful quality control, a clear separation and demarcation of synthetic data from real measurements, and careful interpretation of derived results.

The outcomes of this study lay the groundwork for prospective investigations focused on enhancing \mpli analysis, particularly through the exploration of dedicated cell detection techniques to directly extract cell body instances from \mpli data.
Furthermore, the findings serve as a motivation for gathering additional training data, aiming to refine and extend the application of the virtual \crv staining to a broader range of sections, brains, and species.
While the current model is trained specifically to replicate \crv stainings, the same methodology can in principle be adapted to other staining types, provided appropriate retraining is performed. 
Future research should focus on further investigating its transferability to other datasets, staining protocols, and brains.

  % \clearpage

  % Ethics
  \section*{Ethics Statement}
\label{sec:ethics}
Vervet monkeys used were part of the Vervet Research Colony housed at the Wake Forest School of Medicine.
Our study did not include experimental procedures with living animals.
Brains were obtained when animals were sacrificed to reduce the size of the colony, where they were maintained and sacrificed in accordance with the guidelines of the Wake Forest Institutional Animal Care and Use Committee IACUC \#A11-219 and the AVMA Guidelines for the Euthanasia of Animals.

  % Additional content
  \section*{Data and Code Availability}
\label{sec:data_and_code}

The training pipeline for presented \gramreg and \ganreg models is available on GitHub\footnote{\url{https://github.com/FZJ-INM1-BDA/pli2cells}}.

Code for the online registration \footnote{\url{https://jugit.fz-juelich.de/inm-1/bda/software/data_processing/ffreg}}, data augmentations for \mpli images \footnote{\url{https://jugit.fz-juelich.de/inm-1/bda/software/data_processing/pli-transforms}}, visualization methods for \mpli modalities \footnote{\url{https://jugit.fz-juelich.de/inm-1/bda/software/data_processing/pli-styles}}, as well as additional dependencies \footnote{\url{https://jugit.fz-juelich.de/inm-1/bda/software}} are hosted on our external GitLab server.

ROIs employed for the training and testing of the models in this study, along with a selection of model predictions, are available in our central institutional repository \cite{oberstrass2025dataset}.
In addition, the repository includes whole-slide predictions for test section 559 along with corresponding \mpli modalities.

\section*{Author Contributions}
\label{sec:contribution}
\textbf{Alexander Oberstrass:}
Conceptualization,
Methodology,
Software,
Formal analysis,
Investigation,
Data curation,
Writing - original draft,
Visualization.
\textbf{Esteban Vaca:}
Conceptualization,
Methodology,
Software,
Formal analysis,
Investigation,
Data curation,
Writing - original draft,
Visualization.
\textbf{Eric Upschulte:}
Software,
Formal analysis,
Writing - review \& editing,
Visualization.
\textbf{Meiqi Niu:}
Validation,
Data curation,
Investigation,
Writing - review \& editing.
\textbf{Nicola Palomero-Gallagher:}
Validation,
Investigation,
Writing - review \& editing.
\textbf{David Graessel:}
Validation,
Investigation,
Writing - review \& editing.
\textbf{Christian Schiffer:}
Conceptualization,
Methodology,
Software,
Writing - review \& editing.
\textbf{Markus Axer:}
Conceptualization,
Validation,
Investigation,
Supervision,
Writing - review \& editing,
Project administration,
Funding acquisition.
\textbf{Katrin Amunts:}
Conceptualization,
Validation,
Investigation,
Supervision,
Writing - Review \& Editing,
Resources,
Project administration,
Funding acquisition.
\textbf{Timo Dickscheid:}
Conceptualization,
Methodology,
Supervision,
Writing - original draft,
Writing - Review \& Editing,
Project administration,
Funding acquisition.

\section*{Declaration of Competing Interests}
\label{sec:conflict_of_interest}
The authors declare that the research was conducted in the absence of any commercial or financial relationships that could be construed as a potential conflict of interest.

\section*{Funding}
\label{sec:funding}
This project received funding from the Helmholtz Association’s Initiative and Networking Fund through the Helmholtz International BigBrain Analytics and Learning Laboratory (HIBALL) under the Helmholtz International Lab grant agreement InterLabs-0015, the Helmholtz Association portfolio theme ``Supercomputing and Modeling for the Human Brain'', the Priority Program 2041 (SPP 2041) "Computational Connectomics" of the German Research Foundation (DFG), the Federal Ministry of Education and Research (BMBF) under project number 01GQ1902, the Deutsche Forschungsgemeinschaft (DFG, German Research Foundation; PA 1815/1-1), and the European Union’s Horizon 2020 Research and Innovation Programme, grant agreement 945539 (HBP SGA3), which is now continued in the European Union’s Horizon Europe Programme, grant agreement 101147319 (EBRAINS 2.0 Project). 
Computing time was granted through JARA on the supercomputer JURECA at Jülich Supercomputing Centre (JSC).
Vervet monkey research was supported by the National Institutes of Health under grant agreements R01MH092311 and P40OD010965.

\section*{Acknowledgments}
\label{sec:acknowledgments}
The authors gratefully thank Roger Woods (David Geffen School of Medicine, UCLA, USA) and Karl Zilles (INM-1, Forschungszentrum J\"ulich GmbH, Germany) for their collaboration in the vervet brain project, and the lab team of the INM-1 for preparing the brain sections.
We also sincerely thank the members of the Fiber Architecture and Big Data Analytics groups of the INM-1 for their constructive discussions and thoughtful insights throughout the development of this work.

  \clearpage

  %%%%%%%%%%%%%%%%%%%%%%% References %%%%%%%%%%%%%%%%%%%%%%%%%
  
  \ifx\documenttype\arxiv
      \bibliographystyle{apalike}
      \bibliography{bib/refs_zotero.bib,bib/refs_manual.bib}

@book{gonzalez2008digital,
  title={Digital Image Processing},
  author={Gonzales, Rafael C and Woods, Richard E},
  year={2008},
  publisher={Pearson Education}
}

@incollection{NEURIPS2019_9015,
  title     = {PyTorch: An Imperative Style, High-Performance Deep Learning Library},
  author    = {Paszke, Adam and Gross, Sam and Massa, Francisco and Lerer, Adam and Bradbury, James and Chanan, Gregory and Killeen, Trevor and Lin, Zeming and Gimelshein, Natalia and Antiga, Luca and Desmaison, Alban and Kopf, Andreas and Yang, Edward and DeVito, Zachary and Raison, Martin and Tejani, Alykhan and Chilamkurthy, Sasank and Steiner, Benoit and Fang, Lu and Bai, Junjie and Chintala, Soumith},
  booktitle = {Advances in Neural Information Processing Systems 32},
  pages     = {8024--8035},
  year      = {2019},
  publisher = {Curran Associates, Inc.},
  url       = {http://papers.neurips.cc/paper/9015-pytorch-an-imperative-style-high-performance-deep-learning-library.pdf}
}

@Misc{Yadan2019Hydra,
  author =       {Omry Yadan},
  title =        {Hydra - A framework for elegantly configuring complex applications},
  howpublished = {Github},
  year =         {2019},
  url =          {https://github.com/facebookresearch/hydra}
}

@misc{jirka_borovec_2022_7447212,
  author       = {Jirka Borovec and
                  William Falcon and
                  Akihiro Nitta and
                  Ananya Harsh Jha and
                  otaj and
                  Annika Brundyn and
                  Donal Byrne and
                  Nathan Raw and
                  Shion Matsumoto and
                  Teddy Koker and
                  Brian Ko and
                  Aditya Oke and
                  Sidhant Sundrani and
                  Baruch and
                  Christoph Clement and
                  Clément POIRET and
                  Rohit Gupta and
                  Haswanth Aekula and
                  Adrian Wälchli and
                  Atharva Phatak and
                  Ido Kessler and
                  Jason Wang and
                  JongMok Lee and
                  Shivam Mehta and
                  Zhengyu Yang and
                  Garry O'Donnell and
                  zlapp},
  title        = {Lightning-AI/lightning-bolts: Minor patch release},
  month        = dec,
  year         = 2022,
  howpublished = {Zenodo},
  version      = {0.6.0.post1},
  doi          = {10.5281/zenodo.7447212},
  url          = {https://doi.org/10.5281/zenodo.7447212}
}

@misc{axer2020vervet,
  doi = {10.25493/AFR3-KDK},
  url = {https://kg.ebrains.eu/search/instances/Dataset/79db19fa-41bd-4292-9a33-e0e79dc9a9aa},
  author = {Axer, M. and Gräßel, D. and Palomero-Gallagher, N. and Takemura, H. and Jorgensen, M. J. and Woods, R. and Amunts, K.},
  title = {Images of the nerve fiber architecture at micrometer-resolution in the vervet monkey visual system},
  publisher = {EBRAINS},
  year = {2020},
  copyright = {Creative Commons Attribution Non Commercial Share Alike 4.0 International}
}

@misc{oberstrass2025dataset,
  author = {Oberstrass, Alexander and Vaca-Cerda, Esteban and Upschulte, Eric and Niu, Meiqi and Palomero-Gallagher, Nicola and Graessel, David and Schiffer, Christian and Axer, Markus and Amunts, Katrin and Dickscheid, Timo},
  howpublished = {Jülich DATA},
  title = {{Replication Data for: From Fibers to Cells: Fourier-Based Registration Enables Virtual Cresyl Violet Staining From 3D Polarized Light Imaging}},
  year = {2025},
  version = {V3},
  doi = {10.26165/JUELICH-DATA/ZOBFV7},
  url = {https://doi.org/10.26165/JUELICH-DATA/ZOBFV7}
}

@article{amunts1999,
  title = {Broca's Region Revisited: {{Cytoarchitecture}} and Intersubject Variability},
  shorttitle = {Broca's Region Revisited},
  author = {Amunts, Katrin and Schleicher, Axel and B{\"u}rgel, Uli and Mohlberg, Hartmut and Uylings, Harry B.M. and Zilles, Karl},
  year = {1999},
  journal = {Journal of Comparative Neurology},
  volume = {412},
  number = {2},
  pages = {319--341},
  issn = {1096-9861},
  doi = {10.1002/(SICI)1096-9861(19990920)412:2<319::AID-CNE10>3.0.CO;2-7},
  url = {https://onlinelibrary.wiley.com/doi/abs/10.1002/%28SICI%291096-9861%2819990920%29412%3A2%3C319%3A%3AAID-CNE10%3E3.0.CO%3B2-7},
  urldate = {2025-09-19},
  copyright = {Copyright {\copyright} 1999 Wiley-Liss, Inc.},
  langid = {english}
}

@article{amunts2005,
  title = {Cytoarchitectonic Mapping of the Human Amygdala, Hippocampal Region and Entorhinal Cortex: Intersubject Variability and Probability Maps},
  shorttitle = {Cytoarchitectonic Mapping of the Human Amygdala, Hippocampal Region and Entorhinal Cortex},
  author = {Amunts, K. and Kedo, O. and Kindler, M. and Pieperhoff, P. and Mohlberg, H. and Shah, N.J. and Habel, U. and Schneider, F. and Zilles, K.},
  year = {2005},
  month = dec,
  journal = {Anatomy and Embryology},
  volume = {210},
  number = {5},
  pages = {343--352},
  issn = {1432-0568},
  doi = {10.1007/s00429-005-0025-5},
  url = {https://doi.org/10.1007/s00429-005-0025-5},
  urldate = {2024-04-03},
  langid = {english}
}

@article{amunts2013,
  title = {{{BigBrain}}: {{An Ultrahigh-Resolution 3D Human Brain Model}}},
  shorttitle = {{{BigBrain}}},
  author = {Amunts, Katrin and Lepage, Claude and Borgeat, Louis and Mohlberg, Hartmut and Dickscheid, Timo and Rousseau, Marc-{\'E}tienne and Bludau, Sebastian and Bazin, Pierre-Louis and Lewis, Lindsay B. and {Oros-Peusquens}, Ana-Maria and Shah, Nadim J. and Lippert, Thomas and Zilles, Karl and Evans, Alan C.},
  year = {2013},
  month = jun,
  journal = {Science},
  volume = {340},
  number = {6139},
  pages = {1472--1475},
  publisher = {American Association for the Advancement of Science},
  doi = {10.1126/science.1235381},
  url = {https://www.science.org/doi/full/10.1126/science.1235381},
  urldate = {2024-12-03}
}

@article{amunts2015,
  title = {Architectonic {{Mapping}} of the {{Human Brain}} beyond {{Brodmann}}},
  author = {Amunts, Katrin and Zilles, Karl},
  year = {2015},
  month = dec,
  journal = {Neuron},
  volume = {88},
  number = {6},
  pages = {1086--1107},
  issn = {0896-6273},
  doi = {10.1016/j.neuron.2015.12.001},
  url = {http://www.sciencedirect.com/science/article/pii/S0896627315010727},
  urldate = {2020-12-14},
  langid = {english},
  pmid = {26687219}
}

@article{amunts2020,
  title = {Julich-{{Brain}}: {{A 3D}} Probabilistic Atlas of the Human Brain's Cytoarchitecture},
  shorttitle = {Julich-{{Brain}}},
  author = {Amunts, Katrin and Mohlberg, Hartmut and Bludau, Sebastian and Zilles, Karl},
  year = {2020},
  month = aug,
  journal = {Science},
  volume = {369},
  number = {6506},
  pages = {988--992},
  publisher = {American Association for the Advancement of Science},
  doi = {10.1126/science.abb4588},
  url = {https://www.science.org/doi/full/10.1126/science.abb4588},
  urldate = {2024-12-03}
}

@inproceedings{arganda-carreras2006,
  title = {Consistent and {{Elastic Registration}} of {{Histological Sections Using Vector-Spline Regularization}}},
  booktitle = {Computer {{Vision Approaches}} to {{Medical Image Analysis}}},
  author = {{Arganda-Carreras}, Ignacio and Sorzano, Carlos O. S. and Marabini, Roberto and Carazo, Jos{\'e} Mar{\'i}a and {Ortiz-de-Solorzano}, Carlos and Kybic, Jan},
  editor = {Beichel, Reinhard R. and Sonka, Milan},
  year = {2006},
  series = {Lecture {{Notes}} in {{Computer Science}}},
  pages = {85--95},
  publisher = {Springer},
  address = {Berlin, Heidelberg},
  doi = {10.1007/11889762_8},
  isbn = {978-3-540-46258-3},
  langid = {english}
}

@inproceedings{arjovsky2017a,
  ids = {arjovsky2017},
  title = {Wasserstein {{Generative Adversarial Networks}}},
  booktitle = {Proceedings of the 34th {{International Conference}} on {{Machine Learning}}},
  author = {Arjovsky, Martin and Chintala, Soumith and Bottou, L{\'e}on},
  year = {2017},
  month = jul,
  pages = {214--223},
  publisher = {PMLR},
  issn = {2640-3498},
  url = {https://proceedings.mlr.press/v70/arjovsky17a.html},
  urldate = {2024-12-03},
  langid = {english}
}

@article{axer2011,
  title = {A Novel Approach to the Human Connectome: {{Ultra-high}} Resolution Mapping of Fiber Tracts in the Brain},
  shorttitle = {A Novel Approach to the Human Connectome},
  author = {Axer, Markus and Amunts, Katrin and Gr{\"a}ssel, David and Palm, Christoph and Dammers, J{\"u}rgen and Axer, Hubertus and Pietrzyk, Uwe and Zilles, Karl},
  year = {2011},
  month = jan,
  journal = {NeuroImage},
  volume = {54},
  number = {2},
  pages = {1091--1101},
  issn = {1053-8119},
  doi = {10.1016/j.neuroimage.2010.08.075},
  url = {https://www.sciencedirect.com/science/article/pii/S105381191001178X},
  urldate = {2023-11-17}
}

@article{axer2011a,
  title = {High-{{Resolution Fiber Tract Reconstruction}} in the {{Human Brain}} by {{Means}} of {{Three-Dimensional Polarized Light Imaging}}},
  author = {Axer, Markus and Graessel, David and Kleiner, Melanie and Dammers, Juergen and Dickscheid, Timo and Reckfort, Julia and Huetz, Tim and Eiben, Bjoern and Pietrzyk, Uwe and Zilles, Karl and Amunts, Katrin},
  year = {2011},
  journal = {Frontiers in Neuroinformatics},
  volume = {5},
  issn = {1662-5196},
  doi = {10.3389/fninf.2011.00034},
  url = {https://www.frontiersin.org/articles/10.3389/fninf.2011.00034/full},
  urldate = {2020-06-03},
  langid = {english}
}

@article{axer2022,
  title = {Scale Matters: {{The}} Nested Human Connectome},
  shorttitle = {Scale Matters},
  author = {Axer, Markus and Amunts, Katrin},
  year = {2022},
  month = nov,
  journal = {Science},
  volume = {378},
  number = {6619},
  pages = {500--504},
  publisher = {American Association for the Advancement of Science},
  doi = {10.1126/science.abq2599},
  url = {https://www.science.org/doi/full/10.1126/science.abq2599},
  urldate = {2023-04-28}
}

@article{bielschowsky1904,
  title = {Die {{Silberimpregnations}} Der {{Neurofibrillen}}},
  author = {Bielschowsky, M.},
  year = {1904},
  journal = {J Psychol Neurol},
  volume = {3},
  pages = {169},
  url = {https://cir.nii.ac.jp/crid/1571698601208484608},
  urldate = {2025-04-15}
}

@article{caspers2019,
  title = {Decoding the Microstructural Correlate of Diffusion {{MRI}}},
  author = {Caspers, Svenja and Axer, Markus},
  year = {2019},
  journal = {NMR in Biomedicine},
  volume = {32},
  number = {4},
  pages = {e3779},
  issn = {1099-1492},
  doi = {10.1002/nbm.3779},
  url = {https://onlinelibrary.wiley.com/doi/abs/10.1002/nbm.3779},
  urldate = {2023-10-20},
  copyright = {Copyright {\copyright} 2017 John Wiley \& Sons, Ltd.},
  langid = {english}
}

@article{caspers2022,
  title = {Additional Fiber Orientations in the Sagittal Stratum---Noise or Anatomical Fine Structure?},
  author = {Caspers, Svenja and Axer, Markus and Gr{\"a}{\ss}el, David and Amunts, Katrin},
  year = {2022},
  month = may,
  journal = {Brain Structure and Function},
  volume = {227},
  number = {4},
  pages = {1331--1345},
  issn = {1863-2661},
  doi = {10.1007/s00429-021-02439-w},
  url = {https://doi.org/10.1007/s00429-021-02439-w},
  urldate = {2024-12-03},
  langid = {english}
}

@article{christiansen2018,
  title = {In {{Silico Labeling}}: {{Predicting Fluorescent Labels}} in {{Unlabeled Images}}},
  shorttitle = {In {{Silico Labeling}}},
  author = {Christiansen, Eric M. and Yang, Samuel J. and Ando, D. Michael and Javaherian, Ashkan and Skibinski, Gaia and Lipnick, Scott and Mount, Elliot and O'Neil, Alison and Shah, Kevan and Lee, Alicia K. and Goyal, Piyush and Fedus, William and Poplin, Ryan and Esteva, Andre and Berndl, Marc and Rubin, Lee L. and Nelson, Philip and Finkbeiner, Steven},
  year = {2018},
  month = apr,
  journal = {Cell},
  volume = {173},
  number = {3},
  pages = {792-803.e19},
  publisher = {Elsevier},
  issn = {0092-8674, 1097-4172},
  doi = {10.1016/j.cell.2018.03.040},
  url = {https://www.cell.com/cell/abstract/S0092-8674(18)30364-7},
  urldate = {2024-12-03},
  langid = {english},
  pmid = {29656897}
}

@inproceedings{cohen2018,
  title = {Distribution {{Matching Losses Can Hallucinate Features}} in {{Medical Image Translation}}},
  booktitle = {Medical {{Image Computing}} and {{Computer Assisted Intervention}} -- {{MICCAI}} 2018},
  author = {Cohen, Joseph Paul and Luck, Margaux and Honari, Sina},
  editor = {Frangi, Alejandro F. and Schnabel, Julia A. and Davatzikos, Christos and {Alberola-L{\'o}pez}, Carlos and Fichtinger, Gabor},
  year = {2018},
  series = {Lecture {{Notes}} in {{Computer Science}}},
  pages = {529--536},
  publisher = {Springer International Publishing},
  address = {Cham},
  doi = {10.1007/978-3-030-00928-1_60},
  isbn = {978-3-030-00928-1},
  langid = {english}
}

@article{cross-zamirski2022,
  title = {Label-Free Prediction of Cell Painting from Brightfield Images},
  author = {{Cross-Zamirski}, Jan Oscar and Mouchet, Elizabeth and Williams, Guy and Sch{\"o}nlieb, Carola-Bibiane and Turkki, Riku and Wang, Yinhai},
  year = {2022},
  month = jun,
  journal = {Scientific Reports},
  volume = {12},
  number = {1},
  pages = {10001},
  publisher = {Nature Publishing Group},
  issn = {2045-2322},
  doi = {10.1038/s41598-022-12914-x},
  url = {https://www.nature.com/articles/s41598-022-12914-x},
  urldate = {2024-12-03},
  copyright = {2022 The Author(s)},
  langid = {english}
}

@article{curl2004,
  title = {Quantitative Phase Microscopy: {{A}} New Tool for Investigating the Structure and Function of Unstained Live Cells},
  shorttitle = {Quantitative Phase Microscopy},
  author = {Curl, Claire L and Bellair, Catherine J and Harris, Peter J and Allman, Brendan E and Roberts, Ann and Nugent, Keith A and Delbridge, Lea Md},
  year = {2004},
  month = dec,
  journal = {Clinical and Experimental Pharmacology and Physiology},
  volume = {31},
  number = {12},
  pages = {896--901},
  issn = {0305-1870, 1440-1681},
  doi = {10.1111/j.1440-1681.2004.04100.x},
  url = {https://onlinelibrary.wiley.com/doi/10.1111/j.1440-1681.2004.04100.x},
  urldate = {2024-12-03},
  copyright = {http://onlinelibrary.wiley.com/termsAndConditions\#vor},
  langid = {english}
}

@article{dehaan2021,
  title = {Deep Learning-Based Transformation of {{H}}\&{{E}} Stained Tissues into Special Stains},
  author = {{de Haan}, Kevin and Zhang, Yijie and Zuckerman, Jonathan E. and Liu, Tairan and Sisk, Anthony E. and Diaz, Miguel F. P. and Jen, Kuang-Yu and Nobori, Alexander and Liou, Sofia and Zhang, Sarah and Riahi, Rana and Rivenson, Yair and Wallace, W. Dean and Ozcan, Aydogan},
  year = {2021},
  month = aug,
  journal = {Nature Communications},
  volume = {12},
  number = {1},
  pages = {4884},
  publisher = {Nature Publishing Group},
  issn = {2041-1723},
  doi = {10.1038/s41467-021-25221-2},
  url = {https://www.nature.com/articles/s41467-021-25221-2},
  urldate = {2023-08-22},
  copyright = {2021 The Author(s)},
  langid = {english}
}

@inproceedings{deng2009,
  title = {{{ImageNet}}: {{A}} Large-Scale Hierarchical Image Database},
  shorttitle = {{{ImageNet}}},
  booktitle = {2009 {{IEEE Conference}} on {{Computer Vision}} and {{Pattern Recognition}}},
  author = {Deng, Jia and Dong, Wei and Socher, Richard and Li, Li-Jia and Li, Kai and {Fei-Fei}, Li},
  year = {2009},
  month = jun,
  pages = {248--255},
  issn = {1063-6919},
  doi = {10.1109/CVPR.2009.5206848}
}

@article{ding2016,
  title = {Comprehensive Cellular-Resolution Atlas of the Adult Human Brain},
  author = {Ding, Song-Lin and Royall, Joshua J. and Sunkin, Susan M. and Ng, Lydia and Facer, Benjamin A.C. and Lesnar, Phil and {Guillozet-Bongaarts}, Angie and McMurray, Bergen and Szafer, Aaron and Dolbeare, Tim A. and Stevens, Allison and Tirrell, Lee and Benner, Thomas and Caldejon, Shiella and Dalley, Rachel A. and Dee, Nick and Lau, Christopher and Nyhus, Julie and Reding, Melissa and Riley, Zackery L. and Sandman, David and Shen, Elaine and {van der Kouwe}, Andre and Varjabedian, Ani and Write, Michelle and Zollei, Lilla and Dang, Chinh and Knowles, James A. and Koch, Christof and Phillips, John W. and Sestan, Nenad and Wohnoutka, Paul and Zielke, H. Ronald and Hohmann, John G. and Jones, Allan R. and Bernard, Amy and Hawrylycz, Michael J. and Hof, Patrick R. and Fischl, Bruce and Lein, Ed S.},
  year = {2016},
  journal = {Journal of Comparative Neurology},
  volume = {524},
  number = {16},
  pages = {3127--3481},
  issn = {1096-9861},
  doi = {10.1002/cne.24080},
  url = {https://onlinelibrary.wiley.com/doi/abs/10.1002/cne.24080},
  urldate = {2023-09-18},
  copyright = {Copyright {\copyright} 2016 The Authors The Journal of Comparative Neurology Published by Wiley Periodicals, Inc.},
  langid = {english}
}

@article{dorph-petersen2001,
  title = {Tissue Shrinkage and Unbiased Stereological Estimation of Particle Number and Size},
  author = {{Dorph-Petersen}, K.-A. and Nyengaard, J. R. and Gundersen, H. J. G.},
  year = {2001},
  journal = {Journal of Microscopy},
  volume = {204},
  number = {3},
  pages = {232--246},
  issn = {1365-2818},
  doi = {10.1046/j.1365-2818.2001.00958.x},
  url = {https://onlinelibrary.wiley.com/doi/abs/10.1046/j.1365-2818.2001.00958.x},
  urldate = {2025-09-18},
  langid = {english}
}

@article{fienup1997,
  title = {Invariant Error Metrics for Image Reconstruction},
  author = {Fienup, J. R.},
  year = {1997},
  month = nov,
  journal = {Applied Optics},
  volume = {36},
  number = {32},
  pages = {8352--8357},
  issn = {2155-3165},
  doi = {10.1364/AO.36.008352},
  url = {https://opg.optica.org/ao/abstract.cfm?uri=ao-36-32-8352},
  urldate = {2024-12-03},
  copyright = {{\copyright} 1997 Optical Society of America},
  langid = {english}
}

@inproceedings{gatys2015a,
  title = {Texture {{Synthesis Using Convolutional Neural Networks}}},
  booktitle = {Advances in {{Neural Information Processing Systems}}},
  author = {Gatys, Leon and Ecker, Alexander S and Bethge, Matthias},
  year = {2015},
  volume = {28},
  publisher = {Curran Associates, Inc.},
  url = {https://proceedings.neurips.cc/paper/2015/hash/a5e00132373a7031000fd987a3c9f87b-Abstract.html},
  urldate = {2024-12-03}
}

@inproceedings{goodfellow2014,
  title = {Generative {{Adversarial Nets}}},
  booktitle = {Advances in {{Neural Information Processing Systems}}},
  author = {Goodfellow, Ian and {Pouget-Abadie}, Jean and Mirza, Mehdi and Xu, Bing and {Warde-Farley}, David and Ozair, Sherjil and Courville, Aaron and Bengio, Yoshua},
  year = {2014},
  volume = {27},
  publisher = {Curran Associates, Inc.},
  url = {https://proceedings.neurips.cc/paper_files/paper/2014/hash/5ca3e9b122f61f8f06494c97b1afccf3-Abstract.html},
  urldate = {2024-12-03}
}

@article{guo2020,
  ids = {guo2019},
  title = {Revealing Architectural Order with Quantitative Label-Free Imaging and Deep Learning},
  author = {Guo, Syuan-Ming and Yeh, Li-Hao and Folkesson, Jenny and Ivanov, Ivan E and Krishnan, Anitha P and Keefe, Matthew G and Hashemi, Ezzat and Shin, David and Chhun, Bryant B and Cho, Nathan H and Leonetti, Manuel D and Han, May H and Nowakowski, Tomasz J and Mehta, Shalin B},
  editor = {Forstmann, Birte and Malhotra, Vivek and Van Valen, David},
  year = {2020},
  month = jul,
  journal = {eLife},
  volume = {9},
  pages = {e55502},
  publisher = {eLife Sciences Publications, Ltd},
  issn = {2050-084X},
  doi = {10.7554/eLife.55502},
  url = {https://doi.org/10.7554/eLife.55502},
  urldate = {2024-12-03}
}

@article{hackett2001,
  title = {Architectonic Identification of the Core Region in Auditory Cortex of Macaques, Chimpanzees, and Humans},
  author = {Hackett, Troy A. and Preuss, Todd M. and Kaas, Jon H.},
  year = {2001},
  journal = {Journal of Comparative Neurology},
  volume = {441},
  number = {3},
  pages = {197--222},
  issn = {1096-9861},
  doi = {10.1002/cne.1407},
  url = {https://onlinelibrary.wiley.com/doi/abs/10.1002/cne.1407},
  urldate = {2025-09-26},
  copyright = {Copyright {\copyright} 2001 Wiley-Liss, Inc.},
  langid = {english}
}

@incollection{heilbronner2025,
  title = {The Evolution of the Basal Ganglia},
  booktitle = {Reference {{Module}} in {{Neuroscience}} and {{Biobehavioral Psychology}}},
  author = {Heilbronner, Sarah R. and Mars, Rogier B. and Bijanki, Kelly R. and {Palomero-Gallagher}, Nicola and Tang, Wei},
  year = {2025},
  month = jan,
  publisher = {Elsevier},
  doi = {10.1016/B978-0-443-27380-3.00013-0},
  url = {https://www.sciencedirect.com/science/article/pii/B9780443273803000130},
  urldate = {2025-09-26},
  isbn = {978-0-12-809324-5}
}

@inproceedings{ioffe2015,
  title = {Batch Normalization: Accelerating Deep Network Training by Reducing Internal Covariate Shift},
  shorttitle = {Batch Normalization},
  booktitle = {Proceedings of the 32nd {{International Conference}} on {{International Conference}} on {{Machine Learning}} - {{Volume}} 37},
  author = {Ioffe, Sergey and Szegedy, Christian},
  year = {2015},
  month = jul,
  series = {{{ICML}}'15},
  eprint = {1502.03167},
  pages = {448--456},
  publisher = {JMLR.org},
  address = {Lille, France},
  url = {https://doi.org/10.48550/arXiv.1502.03167},
  urldate = {2024-10-29},
  archiveprefix = {arXiv}
}

@inproceedings{isola2017,
  title = {Image-{{To-Image Translation With Conditional Adversarial Networks}}},
  booktitle = {Proceedings of the {{IEEE Conference}} on {{Computer Vision}} and {{Pattern Recognition}}},
  author = {Isola, Phillip and Zhu, Jun-Yan and Zhou, Tinghui and Efros, Alexei A.},
  year = {2017},
  pages = {1125--1134},
  url = {https://openaccess.thecvf.com/content_cvpr_2017/html/Isola_Image-To-Image_Translation_With_CVPR_2017_paper.html},
  urldate = {2020-12-12}
}

@inproceedings{kingma2017,
  title = {Adam: {{A Method}} for {{Stochastic Optimization}}},
  shorttitle = {Adam},
  booktitle = {3rd {{International Conference}} on {{Learning Representations}}, {{ICLR}}},
  author = {Kingma, Diederik P. and Ba, Jimmy},
  year = {2017},
  eprint = {1412.6980},
  doi = {10.48550/arXiv.1412.6980},
  url = {http://arxiv.org/abs/1412.6980},
  urldate = {2024-10-29},
  archiveprefix = {arXiv}
}

@article{kluver1953,
  title = {A {{Method}} for the {{Combined Staining}} of {{Cells}} and {{Fibers}} in the {{Nervous System}}*},
  author = {Kl{\"u}ver, Heinrich and Barrera, Elizabeth},
  year = {1953},
  month = oct,
  journal = {Journal of Neuropathology \& Experimental Neurology},
  volume = {12},
  number = {4},
  pages = {400--403},
  issn = {0022-3069},
  doi = {10.1097/00005072-195312040-00008},
  url = {https://doi.org/10.1097/00005072-195312040-00008},
  urldate = {2025-04-15}
}

@article{kretschmann1982,
  title = {Different Volume Changes of Cerebral Cortex and White Matter during Histological Preparation},
  author = {Kretschmann, H J and Tafesse, U and Herrmann, A},
  year = {1982},
  month = may,
  journal = {Microscopica acta},
  volume = {86},
  number = {1},
  pages = {13--24},
  issn = {0044-376X},
  langid = {english},
  pmid = {7048029}
}

@inproceedings{kropp2024,
  title = {Denoising {{Diffusion Probabilistic Models}} for {{Image Inpainting}} of {{Cell Distributions}} in {{The Human Brain}}},
  booktitle = {2024 {{IEEE International Symposium}} on {{Biomedical Imaging}} ({{ISBI}})},
  author = {Kropp, Jan-Oliver and Schiffer, Christian and Amunts, Katrin and Dickscheid, Timo},
  year = {2024},
  month = may,
  pages = {1--5},
  issn = {1945-8452},
  doi = {10.1109/ISBI56570.2024.10635384},
  url = {https://ieeexplore.ieee.org/abstract/document/10635384?casa_token=d6nklJ2Egd4AAAAA:dh3OjNel8722Z-X1z5B9az4Jy-uigqv1lMrla7yw9a8vq0Q14dw2TZXyxd9_RWdQeR27CoHoIQ},
  urldate = {2025-03-31}
}

@inproceedings{kuglin1975,
  title = {The Phase Correlation Image Alignment Method},
  booktitle = {Proc. {{International Conference}} on {{Cybernetics}} and {{Society}}, 1975},
  author = {Kuglin, Charles D.},
  year = {1975},
  pages = {163--165}
}

@article{latonen2024,
  title = {Virtual Staining for Histology by Deep Learning},
  author = {Latonen, Leena and Koivukoski, Sonja and Khan, Umair and Ruusuvuori, Pekka},
  year = {2024},
  month = sep,
  journal = {Trends in Biotechnology},
  volume = {42},
  number = {9},
  pages = {1177--1191},
  publisher = {Elsevier},
  issn = {0167-7799, 1879-3096},
  doi = {10.1016/j.tibtech.2024.02.009},
  url = {https://www.cell.com/trends/biotechnology/abstract/S0167-7799(24)00038-6},
  urldate = {2025-08-19},
  langid = {english},
  pmid = {38480025}
}

@article{mirza2014,
  title = {Conditional {{Generative Adversarial Nets}}},
  author = {Mirza, Mehdi and Osindero, Simon},
  year = {2014},
  month = nov,
  journal = {arXiv preprint},
  eprint = {1411.1784},
  doi = {10.48550/arXiv.1411.1784},
  url = {http://arxiv.org/abs/1411.1784},
  urldate = {2024-12-03},
  archiveprefix = {arXiv}
}

@article{nieuwenhuys2013,
  title = {The Myeloarchitectonic Studies on the Human Cerebral Cortex of the {{Vogt-Vogt}} School, and Their Significance for the Interpretation of Functional Neuroimaging Data},
  author = {Nieuwenhuys, Rudolf},
  year = {2013},
  month = mar,
  journal = {Brain Structure \& Function},
  volume = {218},
  number = {2},
  pages = {303--352},
  issn = {1863-2661},
  doi = {10.1007/s00429-012-0460-z},
  langid = {english},
  pmid = {23076375}
}

@article{novotny1977,
  title = {Triple {{Staining}} of {{Normal}} and {{Degenerating Nervous Tissue}}},
  author = {Novotny, G. E. K. and Novotny, E.},
  year = {1977},
  month = jan,
  journal = {Stain Technology},
  volume = {52},
  number = {2},
  pages = {97--99},
  publisher = {Taylor \& Francis},
  issn = {0038-9153},
  doi = {10.3109/10520297709116754},
  url = {https://doi.org/10.3109/10520297709116754},
  urldate = {2025-04-15},
  pmid = {69342}
}

@article{oberstrass2024a,
  ids = {oberstrass2024},
  title = {Self-Supervised Representation Learning for Nerve Fiber Distribution Patterns in {{3D-PLI}}},
  author = {Oberstrass, Alexander and Muenzing, Sascha E.A. and Niu, Meiqi and {Palomero-Gallagher}, Nicola and Schiffer, Christian and Axer, Markus and Amunts, Katrin and Dickscheid, Timo},
  year = {2024},
  month = nov,
  journal = {Imaging Neuroscience},
  volume = {2},
  pages = {1--29},
  issn = {2837-6056},
  doi = {10.1162/imag_a_00351},
  url = {https://doi.org/10.1162/imag_a_00351},
  urldate = {2024-11-14}
}

@article{ojansivu2007,
  title = {Image {{Registration Using Blur-Invariant Phase Correlation}}},
  author = {Ojansivu, Ville and Heikkila, Janne},
  year = {2007},
  month = jul,
  journal = {IEEE Signal Processing Letters},
  volume = {14},
  number = {7},
  pages = {449--452},
  issn = {1558-2361},
  doi = {10.1109/LSP.2006.891338}
}

@article{ounkomol2018,
  title = {Label-Free Prediction of Three-Dimensional Fluorescence Images from Transmitted-Light Microscopy},
  author = {Ounkomol, Chawin and Seshamani, Sharmishtaa and Maleckar, Mary M. and Collman, Forrest and Johnson, Gregory R.},
  year = {2018},
  month = nov,
  journal = {Nature Methods},
  volume = {15},
  number = {11},
  pages = {917--920},
  publisher = {Nature Publishing Group},
  issn = {1548-7105},
  doi = {10.1038/s41592-018-0111-2},
  url = {https://www.nature.com/articles/s41592-018-0111-2},
  urldate = {2024-12-03},
  copyright = {2018 The Author(s), under exclusive licence to Springer Nature America, Inc.},
  langid = {english}
}

@article{park2018,
  title = {Quantitative Phase Imaging in Biomedicine},
  author = {Park, YongKeun and Depeursinge, Christian and Popescu, Gabriel},
  year = {2018},
  month = oct,
  journal = {Nature Photonics},
  volume = {12},
  number = {10},
  pages = {578--589},
  publisher = {Nature Publishing Group},
  issn = {1749-4893},
  doi = {10.1038/s41566-018-0253-x},
  url = {https://www.nature.com/articles/s41566-018-0253-x},
  urldate = {2024-12-03},
  copyright = {2018 Springer Nature Limited},
  langid = {english}
}

@article{pedone2013,
  title = {Blur {{Invariant Translational Image Registration}} for {{N-fold Symmetric Blurs}}},
  author = {Pedone, Matteo and Flusser, Jan and Heikkil{\"a}, Janne},
  year = {2013},
  month = sep,
  journal = {IEEE Transactions on Image Processing},
  volume = {22},
  number = {9},
  pages = {3676--3689},
  issn = {1941-0042},
  doi = {10.1109/TIP.2013.2268972}
}

@article{reddy1996,
  title = {An {{FFT-based}} Technique for Translation, Rotation, and Scale-Invariant Image Registration},
  author = {Reddy, B.S. and Chatterji, B.N.},
  year = {1996},
  month = aug,
  journal = {IEEE Transactions on Image Processing},
  volume = {5},
  number = {8},
  pages = {1266--1271},
  issn = {1941-0042},
  doi = {10.1109/83.506761},
  url = {https://ieeexplore.ieee.org/abstract/document/506761?casa_token=2_WrcRK5JCYAAAAA:fzbvsw7L7XU-f3b1ovfqaaB44oouOipRuxZUpfjNgA59FD34sh-FoIYGiu9ybvYH1rxuNub-sQ},
  urldate = {2024-12-08}
}

@article{rivenson2019,
  title = {{{PhaseStain}}: The Digital Staining of Label-Free Quantitative Phase Microscopy Images Using Deep Learning},
  shorttitle = {{{PhaseStain}}},
  author = {Rivenson, Yair and Liu, Tairan and Wei, Zhensong and Zhang, Yibo and {de Haan}, Kevin and Ozcan, Aydogan},
  year = {2019},
  month = feb,
  journal = {Light: Science \& Applications},
  volume = {8},
  number = {1},
  pages = {23},
  publisher = {Nature Publishing Group},
  issn = {2047-7538},
  doi = {10.1038/s41377-019-0129-y},
  url = {https://www.nature.com/articles/s41377-019-0129-y},
  urldate = {2023-05-09},
  copyright = {2019 The Author(s)},
  langid = {english}
}

@article{rivenson2020,
  title = {Emerging {{Advances}} to {{Transform Histopathology Using Virtual Staining}}},
  author = {Rivenson, Yair and {de Haan}, Kevin and Wallace, W. Dean and Ozcan, Aydogan},
  year = {2020},
  month = aug,
  journal = {BME Frontiers},
  volume = {2020},
  pages = {9647163},
  publisher = {American Association for the Advancement of Science},
  doi = {10.34133/2020/9647163},
  url = {https://spj.science.org/doi/full/10.34133/2020/9647163},
  urldate = {2024-12-03}
}

@inproceedings{ronneberger2015,
  title = {U-{{Net}}: {{Convolutional Networks}} for {{Biomedical Image Segmentation}}},
  shorttitle = {U-{{Net}}},
  booktitle = {Medical {{Image Computing}} and {{Computer-Assisted Intervention}} -- {{MICCAI}} 2015},
  author = {Ronneberger, Olaf and Fischer, Philipp and Brox, Thomas},
  editor = {Navab, Nassir and Hornegger, Joachim and Wells, William M. and Frangi, Alejandro F.},
  year = {2015},
  pages = {234--241},
  publisher = {Springer International Publishing},
  address = {Cham},
  doi = {10.1007/978-3-319-24574-4_28},
  isbn = {978-3-319-24574-4},
  langid = {english}
}

@inproceedings{schiffer2021,
  title = {Contrastive {{Representation Learning For Whole Brain Cytoarchitectonic Mapping In Histological Human Brain Sections}}},
  booktitle = {2021 {{IEEE}} 18th {{International Symposium}} on {{Biomedical Imaging}} ({{ISBI}})},
  author = {Schiffer, Christian and Amunts, Katrin and Harmeling, Stefan and Dickscheid, Timo},
  year = {2021},
  month = apr,
  pages = {603--606},
  issn = {1945-8452},
  doi = {10.1109/ISBI48211.2021.9433986}
}

@article{schiffer2021a,
  title = {Convolutional Neural Networks for Cytoarchitectonic Brain Mapping at Large Scale},
  author = {Schiffer, Christian and Spitzer, Hannah and Kiwitz, Kai and Unger, Nina and Wagstyl, Konrad and Evans, Alan C. and Harmeling, Stefan and Amunts, Katrin and Dickscheid, Timo},
  year = {2021},
  month = oct,
  journal = {NeuroImage},
  volume = {240},
  pages = {118327},
  issn = {1095-9572},
  doi = {10.1016/j.neuroimage.2021.118327},
  langid = {english},
  pmid = {34224853}
}

@article{schleicher2000,
  title = {A Stereological Approach to Human Cortical Architecture: Identification and Delineation of Cortical Areas},
  shorttitle = {A Stereological Approach to Human Cortical Architecture},
  author = {Schleicher, A and Amunts, K and Geyer, S and Kowalski, T and Schormann, T and {Palomero-Gallagher}, N and Zilles, K},
  year = {2000},
  month = oct,
  journal = {Journal of Chemical Neuroanatomy},
  volume = {20},
  number = {1},
  pages = {31--47},
  issn = {0891-0618},
  doi = {10.1016/S0891-0618(00)00076-4},
  url = {https://www.sciencedirect.com/science/article/pii/S0891061800000764},
  urldate = {2021-02-26},
  langid = {english}
}

@inproceedings{schober2015,
  title = {Reference {{Volume Generation}} for {{Subsequent 3D Reconstruction}} of {{Histological Sections}}},
  booktitle = {Bildverarbeitung F{\"u}r Die {{Medizin}} 2015},
  author = {Schober, Martin and Schl{\"o}mer, Philipp and Cremer, Markus and Mohlberg, Hartmut and Huynh, Anh-Minh and Schubert, Nicole and Kirlangic, Mehmet E. and Amunts, Katrin},
  editor = {Handels, Heinz and Deserno, Thomas Martin and Meinzer, Hans-Peter and Tolxdorff, Thomas},
  year = {2015},
  series = {Informatik Aktuell},
  pages = {143--148},
  publisher = {Springer},
  address = {Berlin, Heidelberg},
  doi = {10.1007/978-3-662-46224-9_26},
  isbn = {978-3-662-46224-9},
  langid = {english}
}

@article{sheng1989,
  title = {Fourier-{{Mellin Spatial Filters For Invariant Pattern Recognition}}},
  author = {Sheng, Yunlong},
  year = {1989},
  month = may,
  journal = {Optical Engineering},
  volume = {28},
  number = {5},
  pages = {494--500},
  publisher = {SPIE},
  issn = {0091-3286, 1560-2303},
  doi = {10.1117/12.7976987},
  url = {https://www.spiedigitallibrary.org/journals/optical-engineering/volume-28/issue-5/285494/Fourier-Mellin-Spatial-Filters-For-Invariant-Pattern-Recognition/10.1117/12.7976987.full},
  urldate = {2024-12-08}
}

@article{simonyan2015,
  title = {Very {{Deep Convolutional Networks}} for {{Large-Scale Image Recognition}}},
  author = {Simonyan, Karen and Zisserman, Andrew},
  year = {2015},
  month = apr,
  journal = {arXiv preprint},
  eprint = {1409.1556},
  primaryclass = {cs},
  doi = {10.48550/arXiv.1409.1556},
  url = {http://arxiv.org/abs/1409.1556},
  urldate = {2022-10-14},
  archiveprefix = {arXiv}
}

@inproceedings{spitzer2017,
  title = {Parcellation of Visual Cortex on High-Resolution Histological Brain Sections Using Convolutional Neural Networks},
  booktitle = {2017 {{IEEE}} 14th {{International Symposium}} on {{Biomedical Imaging}} ({{ISBI}} 2017)},
  author = {Spitzer, Hannah and Amunts, Katrin and Harmeling, Stefan and Dickscheid, Timo},
  year = {2017},
  month = apr,
  pages = {920--923},
  issn = {1945-8452},
  doi = {10.1109/ISBI.2017.7950666}
}

@article{stuckner2022,
  title = {Microstructure Segmentation with Deep Learning Encoders Pre-Trained on a Large Microscopy Dataset},
  author = {Stuckner, Joshua and Harder, Bryan and Smith, Timothy M.},
  year = {2022},
  month = sep,
  journal = {npj Computational Materials},
  volume = {8},
  number = {1},
  pages = {1--12},
  publisher = {Nature Publishing Group},
  issn = {2057-3960},
  doi = {10.1038/s41524-022-00878-5},
  url = {https://www.nature.com/articles/s41524-022-00878-5},
  urldate = {2025-02-07},
  copyright = {2022 This is a U.S. Government work and not under copyright protection in the US; foreign copyright protection may apply},
  langid = {english}
}

@article{takemura2020,
  title = {Anatomy of Nerve Fiber Bundles at Micrometer-Resolution in the Vervet Monkey Visual System},
  author = {Takemura, Hiromasa and {Palomero-Gallagher}, Nicola and Axer, Markus and Gr{\"a}{\ss}el, David and Jorgensen, Matthew J and Woods, Roger and Zilles, Karl},
  editor = {Verstynen, Timothy and Behrens, Timothy E and Verstynen, Timothy},
  year = {2020},
  month = aug,
  journal = {eLife},
  volume = {9},
  pages = {e55444},
  publisher = {eLife Sciences Publications, Ltd},
  issn = {2050-084X},
  doi = {10.7554/eLife.55444},
  url = {https://doi.org/10.7554/eLife.55444},
  urldate = {2020-12-17}
}

@article{thornig2021,
  title = {{{JURECA}}: {{Data Centric}} and {{Booster Modules}} Implementing the {{Modular Supercomputing Architecture}} at {{J{\"u}lich Supercomputing Centre}}},
  shorttitle = {{{JURECA}}},
  author = {Th{\"o}rnig, Philipp},
  year = {2021},
  month = oct,
  journal = {Journal of large-scale research facilities JLSRF},
  volume = {7},
  pages = {A182-A182},
  issn = {2364-091X},
  doi = {10.17815/jlsrf-7-182},
  url = {https://jlsrf.org/index.php/lsf/article/view/182},
  urldate = {2023-08-21},
  copyright = {Copyright (c) 2021 Journal of large-scale research facilities JLSRF},
  langid = {english}
}

@article{tong2019,
  title = {Image {{Registration With Fourier-Based Image Correlation}}: {{A Comprehensive Review}} of {{Developments}} and {{Applications}}},
  shorttitle = {Image {{Registration With Fourier-Based Image Correlation}}},
  author = {Tong, Xiaohua and Ye, Zhen and Xu, Yusheng and Gao, Sa and Xie, Huan and Du, Qian and Liu, Shijie and Xu, Xiong and Liu, Sicong and Luan, Kuifeng and Stilla, Uwe},
  year = {2019},
  month = oct,
  journal = {IEEE Journal of Selected Topics in Applied Earth Observations and Remote Sensing},
  volume = {12},
  number = {10},
  pages = {4062--4081},
  issn = {2151-1535},
  doi = {10.1109/JSTARS.2019.2937690}
}

@misc{tran2024,
  title = {Generating Clinical-Grade Pathology Reports from Gigapixel Whole Slide Images with {{HistoGPT}}},
  author = {Tran, Manuel and Schmidle, Paul and Wagner, Sophia J. and Koch, Valentin and Novotny, Brenna and Lupperger, Valerio and Feuchtinger, Annette and B{\"o}hner, Alexander and Kaczmarczyk, Robert and Biedermann, Tilo and Comfere, Nneka I. and Guo, Ruifeng (Ray) and Wang, Chen and Eyerich, Kilian and Braun, Stephan A. and Peng, Tingying and Marr, Carsten},
  year = {2024},
  month = jun,
  pages = {2024.03.15.24304211},
  publisher = {medRxiv},
  doi = {10.1101/2024.03.15.24304211},
  url = {https://www.medrxiv.org/content/10.1101/2024.03.15.24304211v2},
  urldate = {2025-03-31},
  archiveprefix = {medRxiv},
  copyright = {{\copyright} 2024, Posted by Cold Spring Harbor Laboratory. This pre-print is available under a Creative Commons License (Attribution 4.0 International), CC BY 4.0, as described at http://creativecommons.org/licenses/by/4.0/},
  langid = {english}
}

@article{upschulte2022,
  title = {Contour Proposal Networks for Biomedical Instance Segmentation},
  author = {Upschulte, Eric and Harmeling, Stefan and Amunts, Katrin and Dickscheid, Timo},
  year = {2022},
  month = apr,
  journal = {Medical Image Analysis},
  volume = {77},
  pages = {102371},
  issn = {1361-8415},
  doi = {10.1016/j.media.2022.102371},
  url = {https://www.sciencedirect.com/science/article/pii/S136184152200024X},
  urldate = {2023-05-25},
  langid = {english}
}

@inproceedings{upschulte2023,
  title = {Uncertainty-{{Aware Contour Proposal Networks}} for {{Cell Segmentation}} in {{Multi-Modality High-Resolution Microscopy Images}}},
  booktitle = {Proceedings of {{The Cell Segmentation Challenge}} in {{Multi-modality High-Resolution Microscopy Images}}},
  author = {Upschulte, Eric and Harmeling, Stefan and Amunts, Katrin and Dickscheid, Timo},
  year = {2023},
  month = jun,
  pages = {1--12},
  publisher = {PMLR},
  issn = {2640-3498},
  url = {https://proceedings.mlr.press/v212/upschulte23a.html},
  urldate = {2024-04-04},
  langid = {english}
}

@article{wagstyl2020,
  title = {{{BigBrain 3D}} Atlas of Cortical Layers: {{Cortical}} and Laminar Thickness Gradients Diverge in Sensory and Motor Cortices},
  shorttitle = {{{BigBrain 3D}} Atlas of Cortical Layers},
  author = {Wagstyl, Konrad and Larocque, St{\'e}phanie and Cucurull, Guillem and Lepage, Claude and Cohen, Joseph Paul and Bludau, Sebastian and {Palomero-Gallagher}, Nicola and Lewis, Lindsay B. and Funck, Thomas and Spitzer, Hannah and Dickscheid, Timo and Fletcher, Paul C. and Romero, Adriana and Zilles, Karl and Amunts, Katrin and Bengio, Yoshua and Evans, Alan C.},
  year = {2020},
  month = apr,
  journal = {PLOS Biology},
  volume = {18},
  number = {4},
  pages = {e3000678},
  publisher = {Public Library of Science},
  issn = {1545-7885},
  doi = {10.1371/journal.pbio.3000678},
  url = {https://journals.plos.org/plosbiology/article?id=10.1371/journal.pbio.3000678},
  urldate = {2021-09-23},
  langid = {english}
}

@article{wang2004,
  title = {Image Quality Assessment: From Error Visibility to Structural Similarity},
  shorttitle = {Image Quality Assessment},
  author = {Wang, Zhou and Bovik, A.C. and Sheikh, H.R. and Simoncelli, E.P.},
  year = {2004},
  month = apr,
  journal = {IEEE Transactions on Image Processing},
  volume = {13},
  number = {4},
  pages = {600--612},
  issn = {1941-0042},
  doi = {10.1109/TIP.2003.819861}
}

@article{west2013,
  title = {Tissue {{Shrinkage}} and {{Stereological Studies}}},
  author = {West, Mark J.},
  year = {2013},
  month = mar,
  journal = {Cold Spring Harbor Protocols},
  volume = {2013},
  number = {3},
  pages = {pdb.top071860},
  publisher = {Cold Spring Harbor Laboratory Press},
  issn = {1940-3402, 1559-6095},
  doi = {10.1101/pdb.top071860},
  url = {http://cshprotocols.cshlp.org/content/2013/3/pdb.top071860},
  urldate = {2025-09-19},
  langid = {english},
  pmid = {23457338}
}

@article{yang2022,
  title = {Virtual {{Stain Transfer}} in {{Histology}} via {{Cascaded Deep Neural Networks}}},
  author = {Yang, Xilin and Bai, Bijie and Zhang, Yijie and Li, Yuzhu and {de Haan}, Kevin and Liu, Tairan and Ozcan, Aydogan},
  year = {2022},
  month = sep,
  journal = {ACS Photonics},
  volume = {9},
  number = {9},
  pages = {3134--3143},
  publisher = {American Chemical Society},
  doi = {10.1021/acsphotonics.2c00932},
  url = {https://doi.org/10.1021/acsphotonics.2c00932},
  urldate = {2024-12-03}
}

@article{zeineh2017,
  title = {Direct {{Visualization}} and {{Mapping}} of the {{Spatial Course}} of {{Fiber Tracts}} at {{Microscopic Resolution}} in the {{Human Hippocampus}}},
  author = {Zeineh, Michael M. and {Palomero-Gallagher}, Nicola and Axer, Markus and Gr{\"a}{\ss}el, David and Goubran, Maged and Wree, Andreas and Woods, Roger and Amunts, Katrin and Zilles, Karl},
  year = {2017},
  month = mar,
  journal = {Cerebral Cortex (New York, N.Y.: 1991)},
  volume = {27},
  number = {3},
  pages = {1779--1794},
  issn = {1460-2199},
  doi = {10.1093/cercor/bhw010},
  langid = {english},
  pmcid = {PMC5963820},
  pmid = {26874183}
}

@article{zhao2025,
  title = {Hippocampal Architecture Viewed through the Eyes of Methodological Development},
  author = {Zhao, Ling and {Palomero-Gallagher}, Nicola},
  year = {2025},
  month = aug,
  journal = {Anatomical Science International},
  issn = {1447-073X},
  doi = {10.1007/s12565-025-00878-7},
  url = {https://doi.org/10.1007/s12565-025-00878-7},
  urldate = {2025-08-29},
  langid = {english}
}

@inproceedings{zhu2017,
  title = {Unpaired {{Image-To-Image Translation Using Cycle-Consistent Adversarial Networks}}},
  booktitle = {Proceedings of the {{IEEE International Conference}} on {{Computer Vision}}},
  author = {Zhu, Jun-Yan and Park, Taesung and Isola, Phillip and Efros, Alexei A.},
  year = {2017},
  pages = {2223--2232},
  url = {https://openaccess.thecvf.com/content_iccv_2017/html/Zhu_Unpaired_Image-To-Image_Translation_ICCV_2017_paper.html},
  urldate = {2025-05-19}
}

@article{zilles1978,
  title = {Automatic Morphometric Analysis of Retrograde Changes in the Nucleus n. Facialis at Different Ontogenetic Stages in the Rat},
  author = {Zilles, Karl and Schleicher, Axel and Kretschmann, Hans -Joachim},
  year = {1978},
  month = jan,
  journal = {Cell and Tissue Research},
  volume = {190},
  number = {2},
  pages = {285--299},
  issn = {1432-0878},
  doi = {10.1007/BF00218176},
  url = {https://doi.org/10.1007/BF00218176},
  urldate = {2025-05-19},
  langid = {english}
}
  \fi

  \clearpage

  \appendix

  % Appendix
  \section{Appendix}

\begin{table*}[h]
  \centering
  \caption{Specific quantitative results per test ROI, corresponding to averages reported in \cref{tab:eq_effect}. The deviation is reported as standard error over four independent trainings with different random seeds.
  Arrows indicate the direction of better performance ($\uparrow$ higher is better, $\downarrow$ lower is better).
  Best scores per ROI in bold.}
  \label{tab:all_rois}
  \begin{tabular}{llcccc}
    \hline \\[-2ex]
    ROI & Method & MI $\uparrow$ & RMSE $\downarrow$ & SSIM $\uparrow$ & F1 $\uparrow$ \\
    \hline \\[-2ex]
    Motor cortex       & \gram     & 0.061 $\pm$ 0.013 & 33.9 $\pm$ 1.0 & 0.306 $\pm$ 0.024 & 24.8 $\pm$ 3.5 \\
                       & \gramreg  & \textbf{0.168} $\pm$ 0.003 & \textbf{29.2} $\pm$ 0.6 & \textbf{0.446} $\pm$ 0.002 & \textbf{42.9} $\pm$ 0.3 \\
                       & \gan      & 0.117 $\pm$ 0.016 & 34.7 $\pm$ 1.1 & 0.268 $\pm$ 0.010 & 16.8 $\pm$ 1.0 \\
                       & \ganreg   & 0.165 $\pm$ 0.015 & 29.6 $\pm$ 1.0 & 0.394 $\pm$ 0.013 & 36.3 $\pm$ 2.1 \\
    \hline
    Hippocampus        & \gram      & 0.184 $\pm$ 0.015 & 34.5 $\pm$ 0.9 & 0.340 $\pm$ 0.013 & 19.4 $\pm$ 2.3 \\
                       & \gramreg   & \textbf{0.303} $\pm$ 0.003 & \textbf{31.1} $\pm$ 0.5 & \textbf{0.448} $\pm$ 0.001 & \textbf{31.4} $\pm$ 0.4 \\
                       & \gan       & 0.078 $\pm$ 0.016 & 44.3 $\pm$ 3.1 & 0.262 $\pm$ 0.016 & 6.9 $\pm$ 1.3 \\
                       & \ganreg   & 0.151 $\pm$ 0.036 & 32.1 $\pm$ 0.6 & 0.373 $\pm$ 0.027 & 24.9 $\pm$ 1.9 \\
    \hline
    Temporal cortex    & \gram      & 0.077 $\pm$ 0.018 & 33.9 $\pm$ 0.8 & 0.319 $\pm$ 0.040 & 31.7 $\pm$ 4.5 \\
                       & \gramreg   & \textbf{0.236} $\pm$ 0.003 & \textbf{29.1} $\pm$ 0.4 & \textbf{0.497} $\pm$ 0.004 & \textbf{52.7} $\pm$ 0.4 \\
                       & \gan       & 0.051 $\pm$ 0.006 & 43.1 $\pm$ 5.4 & 0.238 $\pm$ 0.019 & 18.3 $\pm$ 2.7 \\
                       & \ganreg   & 0.192 $\pm$ 0.015 & 30.1 $\pm$ 1.5 & 0.460 $\pm$ 0.017 & 49.6 $\pm$ 1.3 \\
    \hline
    Subcortical nuclei & \gram      & 0.122 $\pm$ 0.007 & 32.0 $\pm$ 0.6 & 0.290 $\pm$ 0.016 & 15.7 $\pm$ 2.2 \\
                       & \gramreg   & \textbf{0.190} $\pm$ 0.004 & \textbf{29.2} $\pm$ 0.4 & \textbf{0.388} $\pm$ 0.001 & \textbf{29.4} $\pm$ 0.5 \\
                       & \gan       & 0.138 $\pm$ 0.028 & 33.0 $\pm$ 0.8 & 0.245 $\pm$ 0.013 & 11.4 $\pm$ 0.9 \\
                       & \ganreg   & 0.132 $\pm$ 0.014 & 29.9 $\pm$ 0.7 & 0.330 $\pm$ 0.014 & 20.9 $\pm$ 1.6 \\
    \hline
  \end{tabular}
\end{table*}

\end{document}